\documentclass[aps,prb,twocolumn,showpacs,floatfix,citeautoscript]{revtex4-1}
\pdfoutput=1

\usepackage{amsmath,amssymb,bm}

\allowdisplaybreaks[2]

\usepackage{graphicx}

\newcommand{\Eq}[1]{Eq.~(\ref{#1})}
\newcommand{\Fig}[1]{Fig.~\ref{#1}}

\newcommand{\alphaicancel}{}
\newcommand{\alphaione}{1}
\newcommand{\alphajcancel}{}
\newcommand{\alphajone}{1}

\begin{document}

\title{Corrections to scaling in the critical theory of deconfined criticality} 

\author{Lorenz Bartosch}  
\affiliation{Institut f\"{u}r Theoretische Physik, Universit\"{a}t
  Frankfurt,  60438 Frankfurt, Germany}
\affiliation{Department of Physics, Harvard University, Cambridge, Massachusetts 02138, USA}

\date{July 17, 2013}

\begin{abstract}

Inspired by recent conflicting views on the order of the phase transition from an antiferromagnetic N\'eel state to a valence bond solid, we use the functional renormalization group to study the underlying quantum critical field theory which couples two complex matter fields to a non-compact gauge field. In our functional renormalization group approach we only expand in covariant derivatives of the fields and use a truncation in which the full field dependence of all wave-function renormalization functions is kept. While we do find critical exponents which agree well with some quantum Monte Carlo studies and support the scenario of deconfined criticality, we also obtain an irrelevant eigenvalue of small magnitude, leading to strong corrections to scaling and slow convergence in related numerical studies.

\end{abstract}


\maketitle

\section{Introduction}

The archetypical example for deconfined criticality in condensed matter systems \cite{Senthil04a,*Senthil04b} is the quantum phase transition \cite{SachdevBook} from a N\'eel to valence bond solid (VBS) state in two-dimensional antiferromagnetic spin systems on the square lattice \cite{Read89,*Read90}.
As the N\'eel and VBS states break distinct symmetries, a generic continuous phase transition between these states is not possible within the widely used Landau-Ginzburg-Wilson framework.
However, as was pointed out by Senthil {\em et al.} \cite{Senthil04a,Senthil04b}, oder parameter fluctuations do not necessarily capture the critical fluctuations at a quantum critical point well and their theory of deconfined criticality allows for a continuous quantum phase transition between two different ordered states \cite{Senthil04a,Senthil04b,Levin04,Tesanovic04,Senthil05,Balents05a, *Balents05b,Balents05c}.
In a nutshell, defects in either of the ordered phases carry geometric phase factors and a condensation of these defects induces a competing type of order by destroying the other.
Within the vicinity of the quantum critical point, concepts of scaling and universality are expected to apply, making the theory of deconfined criticality highly predictive.

The critical field theory which is believed to describe the N\'eel to VBS transition on the square or honeycomb lattice is a non-compact CP$^{N-1}$ model in $D = 2+1$ dimensions (with $N=2$ for $S=1/2$ spins).
Relaxing the unit length constraint $\sum_{a=1}^N |\varphi_a|^2 = 1$ of the complex scalar matter fields $\varphi_a$,
this model is also known as the $N$-component superconductor model or the Abelian Higgs model and plays a fundamental role in many different areas of physics. 
Here, $N$ denotes the number of 
fields $\varphi_a$ which interact via a non-compact gauge field $\bm{\mathcal{A}}$. 
In the case of antiferromagnetic spin systems, the N\'eel order parameter is bilinear in the $\varphi_a$ fields,
\begin{equation}
	\label{eq:NeelOrderParameter}
	\bm{N} \propto (\varphi_1^\ast, \varphi_2^\ast) \, \bm{\sigma} \!
\left(
\begin{array}{c}
  \varphi_1   \\
  \varphi_2 \end{array}
\right),
\end{equation}
with $\bm{\sigma}$ the vector of Pauli matrices.
The $\varphi_a$ represent spinor, i.e.\ fractionalized degrees of freedom.
The N\'eel vector $\bm{N}$ is invariant under the local $U(1)$ gauge transformation $\varphi_a \to e^{i \theta} \varphi_a$
and it is this gauge redundancy which necessitates the gauge field.
For general $N$, the action of our field theory is given by
%
\begin{align}
  \label{eq:action}
  & \mathcal{S} [\varphi,\bm{\mathcal{A}}] = 
\int d^D x\, \bigg[ | ( -i {\bm{\nabla}} + \bm{\mathcal{A}}) \varphi_a |^2 \nonumber \\
& {} \qquad \qquad \qquad
+ r |\varphi_a|^2 + \frac{\lambda}{2} (|\varphi_a|^2)^2 + \frac{1}{4 e^2} \mathcal{F}_{\mu \nu}^2 
\bigg] \;.
\end{align}
Here, $\bm{x}$ refers to Euclidean space-time, $\mathcal{F}_{\mu \nu} = \partial_\mu \mathcal{A}_\nu - \partial_\nu \mathcal{A}_\mu$ is the electromagnetic field tensor and a summation over repeated indices is implied. 
For sufficiently large $r$, the fields $\varphi_a$ are gapped and the gauge field $\bm{\mathcal{A}}$ is massless, corresponding to a Coulomb phase with a free photon.
However, as $r$ is reduced below a critical value $r_c$, the condensation of any of the fields $\varphi_a$ leads to a finite mass of the gauge field and the loss of the photon. This is the well-known Higgs transition.
In the context of deconfined criticality, $r$ needs to be fine-tuned to $r_c$ to obtain the scale-invariant critical theory, describing the quantum critical point of the N\'eel to VBS (or spin liquid) transition. 
While monopole operators turn out to be irrelevant at the quantum critical point separating the N\'eel from the VBS state and thus do not have to be included in the above theory, these operators are in fact a relevant perturbation at the Coulomb phase fixed point, turning the spin liquid phase unstable towards the onset of VBS order.

\begin{figure*}[tb]
    \includegraphics*[width=1.4\columnwidth]{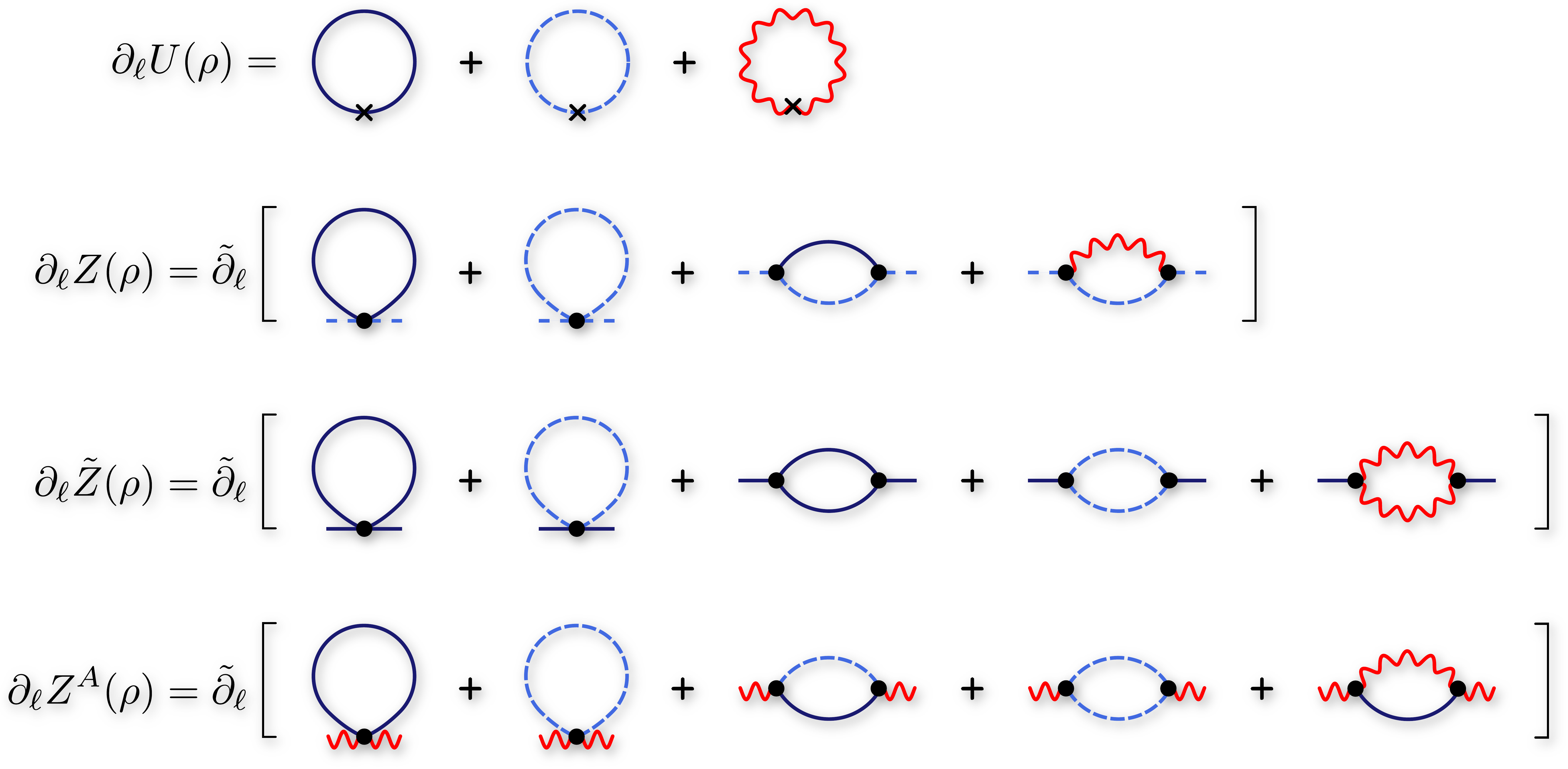}
  \caption{%
(Color online) 
Graphical representation of the flow equations of the effective potential 
$U_\ell(\rho)$ and the wave-function renormalization functions $Z_\ell(\rho)$, $\tilde Z_\ell(\rho)$, 
and $Z_\ell^A (\rho)$, as mathematically described by Eqs.~(\ref{eq:Ubetafunction}) and (\ref{eq:Zbetafunction})--(\ref{eq:ZAbetafunction}) 
in Appendix~\ref{sec:FRGFlowEq}.
While the longitudinal and transverse propagators are depicted by solid and dashed lines, respectively, the gauge field propagator is represented by a wiggly line. The crosses in the flow of $U_\ell(\rho)$ stand for regulator insertions. 
The regulator insertions in the flow equations for the wave-function renormalization functions are generated by the derivative operator $\tilde \partial_\ell$ which only acts on regulator terms. 
}
  \label{fig:Feynman}
\end{figure*}

The gauge theory described by Eq.~(\ref{eq:action}) has a long history with many twists and turns. In a seminal paper from the 70s, Halperin, Lubensky, and Ma \cite {Halperin74} showed within a Wilsonian renormalization group (RG) 
by expanding in $\epsilon = 4 - D$ and extrapolating to $D = 3$ that only for $N \geq 183$ there is a non-trivial fixed point, implying the possibility for a continuous phase transition at a non-zero charge $e^2$. However, no such critical value for $N$ was found within a next-to-leading order $1/N$ expansion \cite{Halperin74,Irkhin96,Kaul08}.
The case of just one complex matter field ($N=1$) was of most interest in those days and it was later shown by Dasgupta and Halperin \cite{Dasgupta81} 
using a duality analysis 
that the theory with $N=1$ lies in the inverted $XY$ universality class.
As a consequence, for $N=1$, the theory defined by Eq.~(\ref{eq:action}) can undergo a continuous phase transition \cite{Dasgupta81,Kiometzis94,Radzihovsky95,Bergerhoff96a,Bergerhoff96b,Folk96,Herbut96}.
In fact, as $D$ is reduced from $D=4$, the critical number of complex field components $N_\mathrm{c}(D)$ above which there exists a stable charged fixed point decreases rapidly \cite{Kolnberger90} and reaches $N_\mathrm{c}(2) = 0$ for $D=2$ \cite{Lawrie83}.

With the advent of ideas of deconfined criticality, the theory with $N=2$ moved center-stage. 
While the most recent studies (of lattice realizations) of Eq.~(\ref{eq:action}) and related spin models such as the sign-problem free $J-Q$ model seem to favor the possibility of a continuous phase transition at a non-zero charge, as hypothesized by the theory of deconfined criticality \cite{Motrunich04,Sandvik07,Nogueira07,Melko08,Motrunich08,Sandvik10,Kaul11,Pujari13,Block13}, other studies are undecided \cite{Harada13} or report weakly first-order phase transitions \cite{Jiang08,Kuklov08,Herland13,Chen13}. Unexpected corrections to scaling were reported by Sandvik, \cite{Sandvik10} and it is our aim here to understand the critical theory and its corrections to scaling from an RG perspective.

\section{Functional Renormalization Group Approach}

In the following, we would like to study the field theory given by Eq.~(\ref{eq:action}) using functional renormalization group (FRG) methods \cite{Berges02,Kopietz10,Delamotte12,Gies12,Metzner12}. 
Following earlier FRG studies \cite{Reuter94b,Bergerhoff96a,Bergerhoff96b,Tetradis97}, we work in the background field formalism which makes it possible to use a gauge-invariant formulation while at the same time fixing a gauge.
The central quantity for which there exists an exact flow equation is the effective average action $\Gamma_\ell [\phi,\bm{A};\bar{\bm{A}}]$, which is explicitly gauge-invariant under a simultaneous gauge transformation of $\phi \equiv \langle \varphi \rangle$ and both gauge fields  $\bm{A} \equiv \langle \bm{\mathcal{A}} \rangle$ and $\bar{\bm{A}}$. Here, $\bar{\bm{A}}$ is the classical background gauge field.
The averages are defined with respect to the action given in Eq.~(\ref{eq:action}) in the presence of sources and regulator terms.
We have parametrized the cutoff scale of the infrared regulator $\Lambda = \Lambda_0 e^{-\ell}$ in terms of the RG time $\ell$.
Continuously removing these regulator terms by increasing $\ell$ from its initial value $\ell = 0$, the effective average action assumes a complicated functional field dependence and becomes the generating functional of irreducible vertices. To make progress, some approximations are necessary. 
Here, we use a derivative expansion in which we expand in (covariant) gradients of $\phi$ and in
$F_{\mu\nu} = \partial_\mu A_\nu -\partial_\nu A_\mu$. 
The ansatz for our truncation reads as
\begin{align}
  \label{eq:AnatzGamma}
 \Gamma_\ell [\phi,\bm{A}] = & 
\int d^dx\, \bigg[ Z_\ell(\rho) | ( -i {\bm{\nabla}} + \bm{A}) \phi_a |^2 \nonumber \\
& {} 
+\frac{1}{2} Y_\ell(\rho) |  {\bm{\nabla}} \rho |^2 + U_\ell (\rho)  + \frac{Z_{\ell}^A (\rho)}{4 e_\ell^2} F_{\mu \nu}^2 
\bigg] ,
\end{align}
where the (gauge-invariant) density $\rho = \phi_a^{\ast} \phi_a$ is a function of $\bm{x}$.
In contrast to previous work, we keep the full functional dependence of the coupling functions $Z_\ell(\rho)$, $Y_\ell(\rho)$, $Z_\ell^A (\rho)$, and $U_\ell(\rho)$. 
Doing so, we also include some momentum-dependence of the four-point vertex and higher-order vertices.
One of the simplest such terms couples to $| \phi_a |^2  | ( -i {\bm{\nabla}} + \bm{A}) \phi_b |^2$.
Let us note that we disentangle the flow of the square of the flowing charge $e_\ell^2$ from the flow of the field-dependent gauge-field renormalization factor $Z_\ell^A (\rho)$ by demanding that for a characteristic density $\rho_{c,\ell}^{\ast}$ we have $Z_\ell^A (\rho_{c,\ell}^{\ast}) = 1$.

The flow equations for the coupling functions are quite involved 
(see Appendix~\ref{sec:FRGFlowEq}),
but do have a simple diagrammatic interpretation, as depicted in Fig.~\ref{fig:Feynman}.
In contrast to perturbation theory, propagators and vertices appearing on the right-hand side of the flow equations are renormalized quantities, involving all powers of interactions in a non-perturbative way and are also non-trivial functions of the density $\rho$.

In order to discuss possible fixed-point properties, it is advantageous to work with rescaled dimensionless variables and coupling functions. To this end, we define
\begin{align}
	\label{eq:scaling:x}
	\tilde{\bm{x}} &= \Lambda \bm{x} , \\
	  \tilde \rho(\tilde {\bm{x}}) &= Z_{\ell}^0 \Lambda^{2-D} \rho(\bm{x}) , \\
  u_\ell(\tilde \rho) &= U_\ell(\rho)/\Lambda^D , \\
  z_\ell(\tilde \rho) & 
  = Z_\ell(\rho)/Z_{\ell}^0 , \\
  \tilde z_\ell(\tilde \rho) &
  =   \left[ Z_\ell(\rho) + \rho Y_\ell (\rho) \right]/Z_{\ell}^0 , \\ 
 z_{\ell}^{A}(\tilde \rho) &= Z_\ell^A(\rho) , \\ 
 \label{eq:flowe2}
  \tilde e_\ell^2 &= \Lambda^{D-4} e_\ell^2 . 
\end{align}
Here, $Z_\ell^0$ is defined as the wave-function renormalization factor evaluated at the characteristic density $\rho_{c,\ell}^{\ast}$, i.e.\ $Z_\ell^0 = Z_\ell (\rho_{c,\ell}^{\ast})$.
For convenience, we choose the corresponding rescaled density $\tilde\rho_{c}^{\ast} = Z_{\ell}^0 \Lambda^{2-D} \rho_{c,\ell}^{\ast}$ to be \mbox{$\ell$-independent} and equal to the position of the minimum of the rescaled effective potential $u_\ell(\tilde\rho)$ at criticality.
It should be noted that by construction $z_\ell (\tilde\rho_{c}^{\ast}) = z_{\ell}^{A} (\tilde\rho_{c}^{\ast}) = 1$. 
The anomalous dimension $\eta_\ell$ of the $\phi$ field and the anomalous dimension $\eta_{A,\ell}$ of the gauge field $\bm{A}$ are related to the flow of $Z_\ell^0$ and 
$1/ \tilde e_\ell^2$ by
\begin{align}
	\label{eq:floweta}
  \eta_\ell & = \partial_\ell \ln Z_{\ell}^0 , \\
  	\label{eq:flowetaA}
  \eta_{A,\ell} & = \partial_\ell \ln \left(1/ \tilde e_\ell^2 \right) .
\end{align}
As we are using the FRG within the background field formalism, both anomalous dimensions are gauge-invariant quantities.
In particular, $\eta$ is the anomalous dimension of a gauge-invariant two-point correlation function of the $\phi$ field at criticality \cite{Sachdev12}.

\section{Results}

Using our above truncation, we obtain a set of partial integro-differential equations (see Appendix~\ref{sec:RescaledFlowEq}) which we turn into a set of ordinary differential equations by choosing a finite mesh for the rescaled density $\tilde \rho$.
At the beginning of our flow in the ultraviolet, $\Gamma_{\ell=0} [\phi,\bm{A}]$ is identical to the bare action given in Eq.~(\ref{eq:action}) and thus completely fixed by the three dimensionless couplings $\tilde r_0$, $\tilde \lambda_0$, and $\tilde e_0^2$, corresponding to $r$, $\lambda$, and $e^2$ in Eq.~(\ref{eq:action}).
Choosing $\tilde \lambda_0$ and $\tilde e_0^2$ to be not too large and positive, it turns out to be always possible to fine-tune 
$\tilde r_0$ such that we approach a non-trivial critical point at a finite charge in the limit of large $\ell$.
\begin{figure}[tb]
  \includegraphics*[width=0.75\columnwidth]{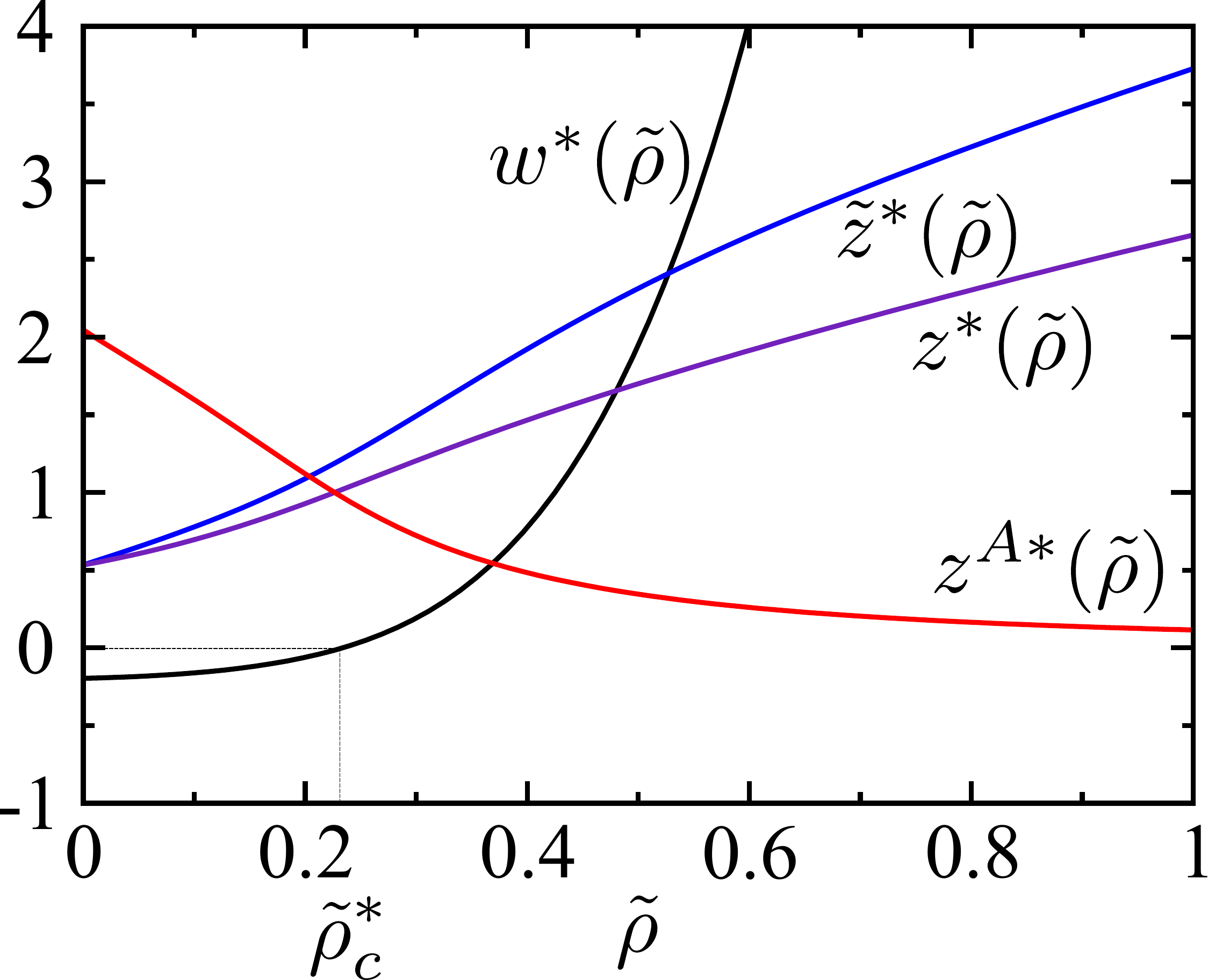}
  \caption{%
(Color online) 
Derivative $w^\ast (\tilde \rho) = d u^\ast (\tilde \rho) / d \tilde \rho$ of the rescaled effective potential  and field-dependent wave-function renormalization functions $z^\ast (\tilde \rho)$, $\tilde z^\ast (\tilde \rho)$, and $z^{A\ast} (\tilde \rho)$ at the charged critical point for $N=2$.}
  \label{fig:fixedpointfunctions}
\end{figure}
At this critical point, our coupling functions $w^\ast (\tilde \rho) \equiv d u^\ast (\tilde \rho)/d\tilde\rho$, $z^\ast (\tilde \rho)$, $\tilde z^\ast (\tilde \rho)$, and $z^{A\ast} (\tilde \rho)$ assume a non-trivial form, as shown graphically for $N=2$ in \Fig{fig:fixedpointfunctions}.

Most interestingly, the wave-function renormalization functions $z^\ast(\tilde\rho)$, $\tilde z^\ast(\tilde\rho)$, and $z^{A\ast}(\tilde\rho)$ are not constant at all and $z^{A\ast}(\tilde\rho)$ even vanishes in the limit of large rescaled densities $\tilde\rho$. However, as we increase the number of complex fields $N$, the wave-function renormalization functions become more and more flat and our results are consistent with previous FRG calculations for large $N$ \cite{Bergerhoff96b}. It should be noted that in previous FRG calculations a first order transition was reported within a derivative expansion for small $N$ \cite{Tetradis97}, and a continuous transition was only found when truncating the effective potential $u_\ell(\tilde\rho)$ at fourth or eighth order in $\phi$ around its minimum \cite{Bergerhoff96a,Bergerhoff96b}. It is therefore reassuring to see that our truncation involving the full functional dependence of both the effective potential and the wave-function renormalization functions leads to a continuous transition for both $N=1$ and $N=2$.

\begin{figure}[b]
  \includegraphics*[width=0.81\columnwidth]{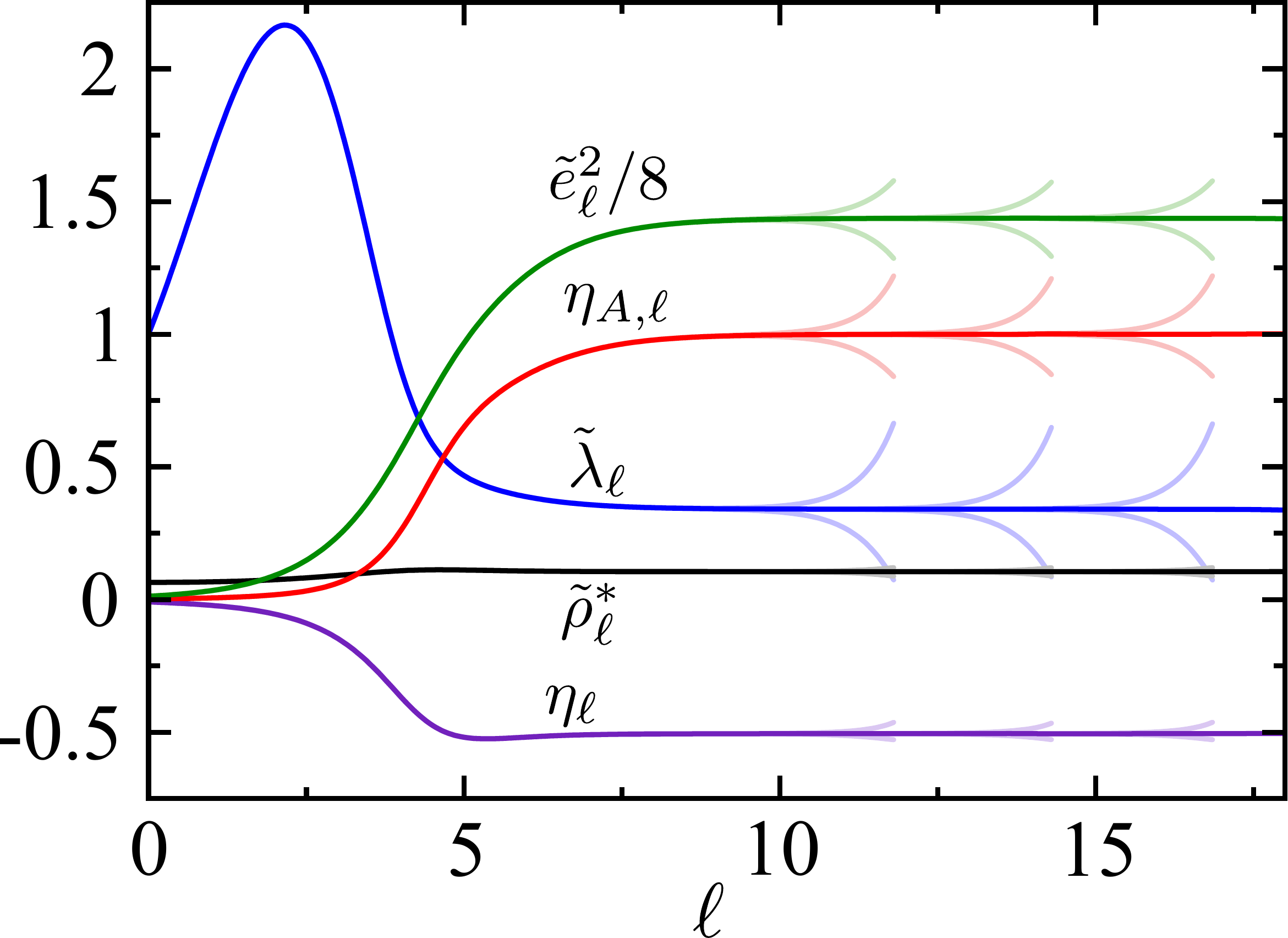}
  \caption{%
(Color online) 
Evolution of the rescaled charge $\tilde e_\ell^2$, the anomalous dimensions of the matter and gauge field $\eta_\ell$ and $\eta_{A,\ell}$, the coupling constant $\tilde\rho_\ell^\ast$ denoting the position of the minimum of the rescaled effective potential, and $\tilde\lambda_\ell$ for {$N=1$}. 
Using a bisection method, we fine-tune the initial value $\tilde\rho_{0}^{\ast}$ (or equivalently $\tilde r_{0} = -\tilde\lambda_0 \tilde\rho_0^{\ast}$) to the value $\tilde\rho_0^{c\ast}$ leading to criticality. 
Choosing $\tilde\rho_0 = \tilde \rho_0^{c\ast} \pm10^{-10},  \tilde \rho_0^{c\ast} \pm 10^{-12},  \tilde \rho_0^{c\ast} \pm 10^{-14}$, we obtain an exponential runaway flow for large $\ell$, as indicated by the light-colored lines.
}
  \label{fig:Evolution1}
\end{figure}

Starting from a small but non-zero charge and fine-tuning $\tilde r_0$ to reach criticality, the flow towards the charged fixed point is shown for $N=1$ in \Fig{fig:Evolution1}. In addition to the flowing charge and the flowing anomalous dimensions of the matter and gauge field, as given by Eqs.~(\ref{eq:flowe2})--(\ref{eq:flowetaA}), we also show the flowing position $\tilde\rho_{\ell}^{\ast}$ of the minimum of $u_\ell(\tilde\rho)$ as well as the projection of the flowing effective potential on its second derivative 
$\tilde\lambda_\ell = d^2 u_\ell / d\tilde\rho^2$
at $\tilde\rho = 0$, corresponding to the coupling constant of a $\phi^4$ term. 
Even though we can in principle get arbitrary close to the critical point by iterated fine-tuning, a relevant operator will finally always drive us away from the critical point.
As the critical exponent $\nu$, which governs the divergence of the correlation length, is just the inverse of the corresponding eigenvalue, it is possible to obtain $\nu$ by simply fitting the deviation of any of the given functions from their critical value to an exponential. 
Doing so, we obtain $\nu \approx 0.56$, which 
is somewhat smaller than the value $\nu_{XY} = 0.67$ 
of the 3D $XY$ model, which by arguments of duality is believed to apply to our field theory \cite{Dasgupta81} with $N=1$.
As can be seen in \Fig{fig:Evolution1}, the anomalous dimension of the matter field flows towards $\eta \approx - 0.5$, while its gauge field counterpart saturates at the exact value \cite{Halperin74} $\eta_A = 1$.
A negative $\eta$ is due to the coupling to the gauge field and was already observed in the $1/N$ expansion \cite{Halperin74} and other studies \cite{Radzihovsky95,Bergerhoff96a,Bergerhoff96b,Folk96,Herbut96}.

\begin{figure}[tb]
  \includegraphics*[width=0.85\columnwidth]{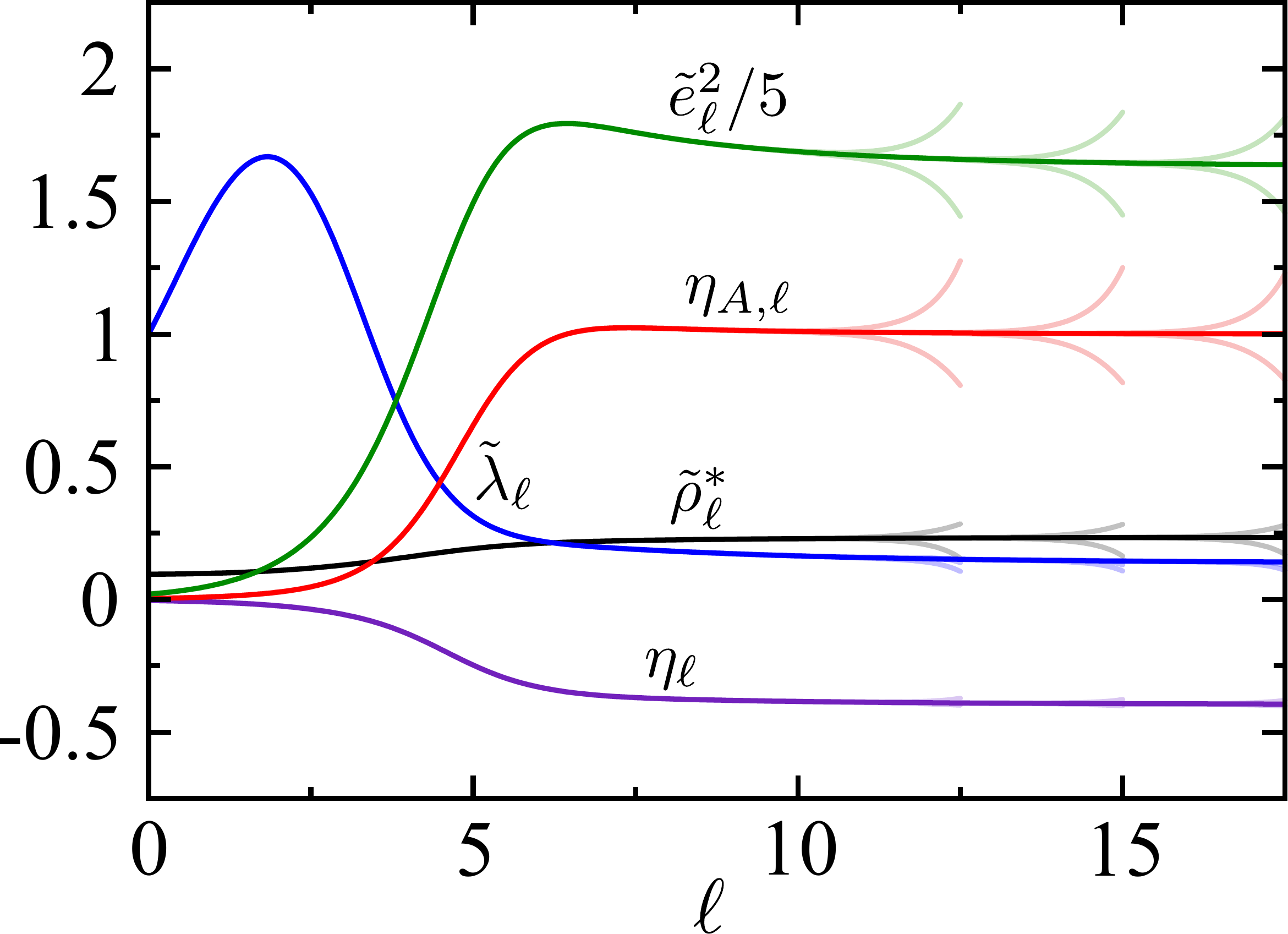}
  \caption{%
(Color online) 
Evolution of $\tilde e_\ell^2$, $\eta_\ell$, $\eta_{A,\ell}$, $\tilde\rho_\ell^\star$, and $\tilde\lambda_\ell$, for initial values chosen as in \Fig{fig:Evolution1}, but for $N=2$, at and near criticality. In contrast to the case $N=1$, the critical point is approached much more slowly.
} 
  \label{fig:Evolution2}
\end{figure}

At first sight, the situation for the $N=2$ case depicted in \Fig{fig:Evolution2} appears to look similar. The critical exponents are in this case given by $\nu \approx 0.56$, $\eta \approx -0.39$, and $\eta_A = 1$.
Notice, however, that the critical point is approached much more slowly than for $N=1$. Quantitatively, we can fit the deviation of the functions depicted in Figs.~\ref{fig:Evolution1} and \ref{fig:Evolution2} from their critical values in the region where the critical point is approached to an exponential and thereby obtain the irrelevant eigenvalue with the smallest magnitude. 
The dominant correction-to-scaling exponent $\omega$ is then defined as the absolute value of this eigenvalue.
In fact, while for $N=1$ we obtain $\omega \approx 1.1$, 
for $N=2$ we obtain the much smaller value $\omega \approx 0.3$. 
Corrections to scaling are therefore much stronger for $N=2$ then for $N=1$ and a considerably longer RG time is needed to reach criticality.

As we basically do obtain the same critical exponent $\nu$ for $N = 1$ and $2$, it is likely that the true $N = 2$ value for $\nu$ is also close to $\nu = 0.67$ \cite{Footnote}. We would like to note that even though our value for $\nu$ might come out a bit too small, an even smaller value, $\nu = 0.51$, was reported in a recent quantum Monte Carlo calculation of the $t$--$Q$ model on the honeycomb lattice which is believed to lie in the same universality class \cite{Pujari13}.
Ignoring vertex corrections, the spin--spin correlation function $\langle \bm{N} (\bm{x}) \cdot \bm{N} (\bm{0})  \rangle \propto 1/|\bm{x}|^{D-2+\eta_N}$ with the N\'eel field given by \Eq{eq:NeelOrderParameter} is proportional to the square of the gauge-invariant two-point correlator of the sclar field $\varphi$ which itself is proportional to $1/|\bm{x}|^{D-2+\eta}$. 
As in a standard $1/N$ calculation with a gauge-fixing parameter the vertex contribution to the anomalous dimension has a gauge-dependent sign and turns out to be relatively small even when extrapolating the $1/N$ result down to $N = 2$ \cite{Kaul08}, the resulting expression $\eta_{N} = D-2 + 2 \eta$ likely represents a good approximation for the anomalous dimension of the N\'eel field.
In fact, our result $\eta_{N} = 0.22$ agrees well with Monte Carlo calculations which predict values in the range $0.2$ -- $0.4$ \cite{Footnote}.


\section{Conclusions}

In summary, we have used functional renormalization group methods to study the critical field theory of deconfined criticality, as emerging in the continuous N\'eel to VBS transition in a class of two-dimensional spin systems. In contrast to previous functional renormalization group studies of the same field theory, we have used a truncation in which the complete field dependence of all wave-function renormalization functions is kept and an expansion only in gradient terms is made. Our results are consistent with some recent (quantum) Monte Carlo calculations and support the scenario of deconfined criticality. However, in contrast to lattice calculations where extrapolation to the infinite system size is an issue and critical properties of the underlying continuum field theory are difficult to address, using functional renormalization group methods it is possible to directly work in the continuum and with an infinite system. In particular, we can start with a very small charge and still reach a critical point for which we determine critical exponents and have also access to irrelevant eigenvalues. Interestingly, the dominant correction-to-scaling exponent is much smaller for $N=2$ than for $N=1$ which explains slow convergence in related numerical studies with system size. 

\acknowledgments

We would like to thank Stephen Powell, Subir Sachdev, and Philipp Strack for helpful discussions.
This work was supported by the DFG via the research group FOR\,723.


\appendix

\section{FRG flow equations}
\label{sec:FRGFlowEq}

The central object of our functional renormalization group (FRG) study is the effective average action which is defined as the Legendre transform of the generating functional of connected Green functions \cite{Berges02,Kopietz10,Delamotte12,Gies12,Metzner12}. 
Roughly speaking, a regulator $R_\ell$ is introduced to give all fluctuations with momenta $|\bm{q}| \lesssim \Lambda = \Lambda_0 e^{-\ell}$ an artificial mass.
The effective average action then contains only quantum fluctuations with momenta larger then the infrared cutoff $\Lambda$.
As the regulator is removed during the evolution of the flow, 
the effective average action turns into the generating functional of one-particle irreducible Green functions.

In this work,
we use the FRG for the Legendre effective average action in the background field formalism \cite{Reuter94b}. 
The background field formalism has the advantage of allowing for a gauge-invariant formulation while at the same time fixing a gauge and including a regulator. 
This, however, comes at the price of having to split the dynamic gauge field $\bm{\mathcal{A}}$ into a non-quantized background field $\bar{\bm{A}}$ and a fluctuating field $\bm{a}$ by setting 
$\bm{\mathcal{A}} = \bar{\bm{A}} + \bm{a}$.
Using a gauge-fixing condition which only involves the combination $\bm{\mathcal{A}} - \bar{\bm{A}}$ and coupling only this combination to external sources, one can derive an effective average action $\Gamma_\ell [\phi,\bm{A};\bar{\bm{A}}]$ via a Legendre transformation which is explicitly gauge-invariant under a simultaneous gauge transformation of $\phi = \langle \varphi \rangle$ and both gauge fields  $\bm{A} = \langle \bm{\mathcal{A}} \rangle$ and $\bar{\bm{A}}$. 
The averages here are averages with respect to the action given in Eq.~(\ref{eq:action}) 
in the presence of sources and regulator terms.
While it is possible to write an exact flow equation for 
$\Gamma_\ell [\phi,\bm{A};\bar{\bm{A}}]$, 
one finally would like to eliminate the background field $\bar{\bm{A}}$ by identifying it with $\bm{A}$.
The main problem in doing so stems from the fact that the functional derivatives with respect to $\bar{\bm{A}}$ and $\bm{A}$ do not coincide.
Partially, this difference can be absorbed by introducing the gauge-invariant normalization factor $\mathcal{C}_\ell [\phi,\bm{A}]$, which vanishes for $\bm{A}=0$ and in both limits $\Lambda \to 0$ and $\Lambda \to \infty$. This term is discussed in detail in Ref.~\onlinecite{Reuter94b}.
Using this strategy, Reuter and Wetterich define a gauge-invariant effective average action \cite{Reuter94b} $\Gamma_\ell [\phi,\bm{A}]$, satisfying the approximate flow equation
\begin{align}
  \label{eq:FlowEquation}
  & \frac{\partial}{\partial \ell} \Gamma_\ell [\phi,\bm{A}] \nonumber \\
& \quad = \frac{1}{2} \text{Tr} 
  \left[ \frac{\partial}{\partial \ell} \bm{\mathcal{R}}_\ell[\bm{A}]\left(\bm{\Gamma}_\ell^{(2)}[\phi,\bm{A}] + 
      \bm{\Gamma}_\text{gf}^{(2)} + \bm{\mathcal{R}}_\ell[\bm{A}]\right)^{-1} \right] \nonumber \\
& {} \qquad  + 
  \frac{\partial}{\partial \ell} \mathcal{C}_\ell [\phi,\bm{A}] .
\end{align}
Here, $\bm{\Gamma}_\ell^{(2)}[\phi,\bm{A}]$ is the matrix of second functional derivatives of $\Gamma_\ell [\phi,\bm{A}]$ with respect to the fields, $\bm{\Gamma}_\text{gf}^{(2)}$ is the corresponding matrix of the gauge-fixing potential, and $\bm{\mathcal{R}}_\ell[\bm{A}]$ is the regulator in matrix form.
In addition to an integration over momentum space, the trace involves a sum over all internal degrees of freedom, i.e.
\begin{equation}
  \label{eq:Trace}
   \text{Tr}\,[\dots]= \int \frac{d^D k}{(2\pi)^D} \sum \,[\dots] .
\end{equation}
For the complex fields $\phi = (\phi_1 \dots \phi_N)$, the sum runs over the $N$ components $a = 1 \dots N$ and contains also the label $i=1,2$, distinguishing the real and imaginary part of $\phi_a = (\phi_{a,1} + i \phi_{a,2})/\sqrt{2}$.
As concerns the gauge field $\bm{A} = (A_1 \dots A_D)$, the sum just runs over $\mu = 1 \dots D$.
Finally we will be interested in the physical case $D = 2 + 1 = 3$.
For the gauge-fixing potential, we follow previous work \cite{Reuter94b,Bergerhoff96a,Bergerhoff96b} and choose
\begin{equation}
  \label{eq:Gammagf}
  \Gamma_\text{gf} [\bm{A};\bar{\bm{A}}] = \frac{1}{2 \alpha} \int d^D r\,
  \left[\bm{\nabla} \cdot (\bm{A} - \bar{\bm{A}}) \right]^2 .
\end{equation}
Taking the limit $\alpha \to 0$ amounts to the background Landau gauge
\begin{equation}
  \label{eq:gaugecond}
  \bm{\nabla} \cdot (\bm{A} - \bar{\bm{A}})  = 0 .
\end{equation}
Of course, Eq.~(\ref{eq:FlowEquation}) involves an infinite set of operators and cannot be solved exactly. To make progress, we use a derivative expansion in which we keep the full functional dependence of the functions $Z_\ell(\rho)$, $Y_\ell(\rho)$, $Z_\ell^A (\rho)$, and $U_\ell(\rho)$ entering our ansatz for the effective average action, as given in \Eq{eq:action} of the paper.
In order to determine the flow of these coupling functions, we expand the right-hand side of Eq.~(\ref{eq:FlowEquation}) in deviations from the space-independent field configuration $\phi_{a,i}(\bm{r}) = \phi_{a,i}$, $\bm{A}(\bm{r}) = 0$. This is facilitated by the inherent 
$U(1) \times SU(N)$ symmetry of the effective average action which allows us to choose $\phi_{1,1} = \sqrt{2 \rho}$ and $\phi_{a,i} = 0$ for all other components. 
Projecting the flow of $\Gamma_\ell [\phi,\bm{A}]$ onto the flow of $U_\ell (\rho)$ by considering a space-independent field configuration, we thereby obtain with $\int_{\bm{k}}  \equiv \int {d^D k}/{(2\pi)^D}$ 
\begin{align}
  & \partial_\ell U_\ell (\rho) = 
\frac{1}{2} \int_{\bm{k}} 
 \Big( \partial_\ell R_\ell^\phi (k^2)
  \big[ G_\ell^L(k^2;\rho) \nonumber \\
 & \ + (2N-1)G_\ell^T(k^2;\rho) \big]
 + (D-1) \partial_\ell R_\ell^A (k^2) G_\ell^A(k^2;\rho) \Big)
 ,
  \label{eq:Ubetafunction}
\end{align}
where 
\begin{align}
  & G_\ell^L (k^2;\rho) \nonumber \\ 
  & \ =  
\frac{1}{\left[ Z_\ell (\rho) + \rho Y_\ell (\rho) \right] k^2 + U_\ell' (\rho) + 2 \rho U_\ell''(\rho) + R_\ell^\phi (k^2)}  , \\[1mm]
  & G_\ell^T (k^2;\rho) =  
\frac{1}{Z_\ell (\rho) k^2 + U_\ell'(\rho) + R_\ell^\phi  (k^2)} 
\end{align}
are the longitudinal and transverse propagators for the given field configuration and $G_\ell^A(k^2;\rho)$ is defined in terms of the gauge-field propagator $\bar G_{\ell;\mu \nu}^A (k^2;\rho)$ by
\begin{align}
  \label{eq:GA}
  \bar G_{\ell;\mu \nu}^A (k^2;\rho) & = \left(\delta_{\mu,\nu} - k_\mu k_\nu/k^2\right) G_\ell^A (k^2;\rho) \nonumber \\
 & =
\frac{\delta_{\mu,\nu} - k_\mu k_\nu/k^2}{(Z_\ell^A (\rho)/e_\ell^2) k^2 + 2\rho Z_\ell(\rho) + R_\ell^A  (k^2)} .
\end{align}
The regulator functions $R_\ell^\phi  (k^2)$ and $R_\ell^A  (k^2)$ will be specified below.
Following a standard recipe \cite{Berges02,Kopietz10,Delamotte12,Gies12,Metzner12}, we can also derive flow equations for the wave-function renormalization factors $Z_\ell(\rho)$, $\tilde Z_\ell(\rho) = Z_\ell(\rho) + \rho Y_\ell (\rho)$, and $Z_\ell^A (\rho)/e_\ell^2$:
\begin{widetext}
\begin{align}
 \partial_\ell Z_\ell (\rho)  =  \frac{\partial}{\partial p^2}\Bigg[ &  
    \frac{1}{2} \tilde \partial_\ell  
    \int_{\bm{k}} \Big(
    \Gamma_\ell^{T,T,L,L} (-\bm{p},\bm{p},-\bm{k},\bm{k};\rho) G_\ell^L (k^2;\rho) \nonumber \\ 
 &  \quad {} + \Gamma_\ell^{T,T,T,T} (-\bm{p},\bm{p},-\bm{k},\bm{k};\rho)
    G_\ell^T (k^2;\rho)
    + (2N-2) \Gamma_\ell^{T,T,T',T'} (-\bm{p},\bm{p},-\bm{k},\bm{k};\rho)
    G_\ell^T (k^2;\rho) \Big) \nonumber \\ 
  {}  + &
    \frac{1}{2} \tilde \partial_\ell  
    \int_{\bm{k}}
    \Gamma_{\ell;\mu \nu}^{A,A,T,T} (-\bm{k},\bm{k},-\bm{p},\bm{p};\rho)
    \bar G_{\ell;\mu \nu}^A (k^2;\rho)  \nonumber \\ 
  {}  - &
  \tilde \partial_\ell  
    \int_{\bm{k}} \Gamma_\ell^{T,T,L} (-\bm{p},-\bm{k},\bm{k} + \bm{p};\rho)
    \Gamma_\ell^{T,T,L} (\bm{p},\bm{k},-(\bm{k} + \bm{p});\rho)
G_\ell^T (k^2;\rho) G_\ell^L ((\bm{k}+\bm{p})^2;\rho)  \nonumber \\ 
  {}  - &
  \tilde \partial_\ell  
    \int_{\bm{k}} \Gamma_{\ell;\mu}^{A,T,T} (\bm{k} + \bm{p},-\bm{p},-\bm{k};\rho)
    \Gamma_{\ell;\nu}^{A,T,T} (-(\bm{k} + \bm{p}),\bm{p},\bm{k};\rho)
G_\ell^T (k^2;\rho) \bar G_{\ell;\mu \nu}^A ((\bm{k}+\bm{p})^2;\rho) \Bigg]
 ,
  \label{eq:Zbetafunction}
\end{align}

\begin{align}
  \partial_\ell \tilde Z_\ell (\rho)   =  \frac{\partial}{\partial p^2}\Bigg[ &  
    \frac{1}{2} \tilde \partial_\ell  
    \int_{\bm{k}} \Big(
    \Gamma_\ell^{L,L,L,L} (-\bm{p},\bm{p},-\bm{k},\bm{k};\rho) G_\ell^L (k^2;\rho) 
   + (2N-1) \Gamma_\ell^{L,L,T,T} (-\bm{p},\bm{p},-\bm{k},\bm{k};\rho)
    G_\ell^T (k^2;\rho) \Big) \nonumber \\ 
  {}  + &
    \frac{1}{2} \tilde \partial_\ell  
    \int_{\bm{k}}
    \Gamma_{\ell;\mu \nu}^{A,A,L,L} (-\bm{k},\bm{k},-\bm{p},\bm{p};\rho)
    \bar G_{\ell;\mu \nu}^A (k^2;\rho)  \nonumber \\ 
  {}  - &
     \frac{1}{2} 
  \tilde \partial_\ell 
    \int_{\bm{k}} \Big( \Gamma_\ell^{L,L,L} (-\bm{p},-\bm{k},\bm{k} + \bm{p};\rho)
    \Gamma_\ell^{L,L,L} (\bm{p},\bm{k},-(\bm{k} + \bm{p});\rho)
G_\ell^L (k^2;\rho) G_\ell^L ((\bm{k}+\bm{p})^2;\rho) \nonumber \\
 &  \quad {} + \Gamma_\ell^{L,T,T} (-\bm{p},-\bm{k},\bm{k} + \bm{p};\rho)
    \Gamma_\ell^{L,T,T} (\bm{p},\bm{k},-(\bm{k} + \bm{p});\rho)
G_\ell^T (k^2;\rho) G_\ell^T ((\bm{k}+\bm{p})^2;\rho) \Big)
 \nonumber \\ 
  {}  - &
  \tilde \partial_\ell  
    \int_{\bm{k}} \Gamma_{\ell;\mu}^{A,L,L} (\bm{k} + \bm{p},-\bm{p},-\bm{k};\rho)
    \Gamma_{\ell;\nu}^{A,L,L} (-(\bm{k} + \bm{p}),\bm{p},\bm{k};\rho)
G_\ell^L (k^2;\rho) \bar G_{\ell;\mu \nu}^A ((\bm{k}+\bm{p})^2;\rho)
 \nonumber \\ 
  {}  - &
     \frac{1}{2} 
  \tilde \partial_\ell  
    \int_{\bm{k}} \Gamma_{\ell;\mu \mu'}^{A,A,L} (-\bm{k},\bm{k} + \bm{p},-\bm{p};\rho)
    \Gamma_{\ell;\nu \nu'}^{A,A,L} (\bm{k},-(\bm{k} + \bm{p}),\bm{p};\rho)
\bar G_{\ell;\mu' \nu'}^A (k^2;\rho) \bar G_{\ell;\mu \nu}^A ((\bm{k}+\bm{p})^2;\rho) \Bigg]
 ,
  \label{eq:tildeZbetafunction}
\end{align}
\begin{align}
  \partial_\ell \left(\frac{Z_\ell^A(\rho )}{e_\ell^2}\right) =& \frac{\partial}{\partial  (p^2 \delta_{\mu,\nu} - p_\mu p_\nu)}\Bigg[
    \frac{1}{2} \tilde \partial_\ell  
    \int_{\bm{k}} \Big(
    \Gamma_{\ell;\mu \nu}^{A,A,L,L} (-\bm{p},\bm{p},-\bm{k},\bm{k};\rho) G_\ell^L (k^2;\rho) 
 \nonumber \\  &  \qquad  \qquad  \qquad  \qquad  \qquad  \qquad {} 
   + (2N-1) \Gamma_{\ell;\mu \nu}^{A,A,T,T} (-\bm{p},\bm{p},-\bm{k},\bm{k};\rho) G_\ell^T (k^2;\rho) \Big) \nonumber \\
   &  \qquad {}  +  \tilde \partial_\ell  
    \int_{\bm{k}} \Big( \left[ \Gamma_{\ell;\mu}^{A,L,T} (-\bm{p},-\bm{k},\bm{k} + \bm{p};\rho) - i (2 k_\mu + p_\mu) R_\ell^{\phi \prime} (0) \right] \nonumber \\ 
 &   \qquad  \qquad \qquad  \qquad  {} \times \left[
    \Gamma_{\ell;\nu}^{A,L,T} (\bm{p},\bm{k},-(\bm{k} + \bm{p});\rho) 
+ i (2 k_\mu + p_\mu) R_\ell^{\phi \prime} (0) \right]
G_\ell^L (k^2;\rho) G_{\ell}^T ((\bm{k}+\bm{p})^2;\rho) \nonumber \\ 
 &   \qquad  \qquad   \qquad  {} + (N-1) \left[ \Gamma_{\ell;\mu}^{A,T,T} (-\bm{p},-\bm{k},\bm{k} + \bm{p};\rho)  - i (2 k_\mu + p_\mu) R_\ell^{\phi \prime} (0) \right] \nonumber \\ 
 &   \qquad  \qquad \qquad  \qquad  {} \times \left[
    \Gamma_{\ell;\nu}^{A,T,T} (\bm{p},\bm{k},-(\bm{k} + \bm{p});\rho)
+ i (2 k_\mu + p_\mu) R_\ell^{\phi \prime} (0) \right]
G_\ell^T (k^2;\rho) G_{\ell}^T ((\bm{k}+\bm{p})^2;\rho) \Big)
 \nonumber \\ 
   &  \qquad {} -
  \tilde \partial_\ell  
    \int_{\bm{k}} \Gamma_{\ell;\mu \mu'}^{A,A,L} (-\bm{p},-\bm{k},\bm{k} + \bm{p};\rho)
    \Gamma_{\ell;\nu \nu'}^{A,A,L} (\bm{p},\bm{k},-(\bm{k} + \bm{p});\rho)
\bar G_{\ell;\mu' \nu'}^A (k^2;\rho) G_{\ell}^L ((\bm{k}+\bm{p})^2;\rho) \Bigg]
 \nonumber \\ 
   &  \qquad {} 
+   \partial_\ell \left(\frac{Z_\ell^{\mathcal{C}} (\rho )}{e_\ell^2}\right) 
 .
  \label{eq:ZAbetafunction}
\end{align}
\end{widetext}
Within a derivative expansion, it is customary to evaluate these wave-function renormalization factors in the limit $p^2 \to 0$.
In contrast to $\partial_\ell$, the partial derivative $\tilde\partial_\ell$ appearing above acts only on regulator terms to its right-hand side, e.g.
when acting on a propagator $G_\ell^i (k^2;\rho)$, this propagator is replaced by the corresponding single-scale propagator
\begin{equation}
  \label{eq:singlescale}
  \tilde\partial_\ell G_\ell^i (k^2;\rho) = - \left[G_\ell^i (k^2;\rho) \right]^2  \partial_\ell R_\ell^i (k^2) .
\end{equation}
Note that due to the residual gauge field dependence of the regulator $\bm{\mathcal{R}}_\ell[\bm{A}]$, the first derivative 
\begin{equation}
  \label{eq:Rderivative}
  \left. R_\ell^{\phi \prime} (0) = \frac{d}{d(k^2)} R_\ell^\phi (k^2) \right|_{k^2 = 0}
\end{equation}
appears on the 
right-hand-side of Eq.~(\ref{eq:ZAbetafunction}). In principle, there is also a term containing the second derivative of $R_\ell^\phi (k^2)$, but this term vanishes for the Litim cutoff which we will use in Appendix~\ref{sec:RescaledFlowEq}.
The Feynman diagrams corresponding to the right-hand-sides of the flow equations of the effective potential $U_\ell(\rho)$ and the wave-function renormalization factors $Z_\ell(\rho)$, $\tilde Z_\ell(\rho)$, 
and $Z_\ell^A (\rho)/e_\ell^2$, as given by Eqs.~(\ref{eq:Ubetafunction}) and (\ref{eq:Zbetafunction})--(\ref{eq:ZAbetafunction}), are depicted in Fig.~\ref{fig:Feynman}. 
As the number of transverse fields entering a given vertex always has to be even, there is no term in the flow of $Z_\ell(\rho)$ corresponding to the last term on the right hand side of Eq.~(\ref{eq:tildeZbetafunction}).
We note that the last term in Eq.~(\ref{eq:ZAbetafunction}) results from the normalization factor $\mathcal{C}_\ell [\phi,\bm{A}]$ and is discussed in detail in Ref.~\onlinecite{Reuter94b}.
Taking derivatives of the effective average action with respect to the fields, we obtain all vertices appearing above,
\begin{widetext}
\begin{align}
 \Gamma_\ell^{L,L,L,L} (-\bm{p},\bm{p},-\bm{k},\bm{k};\rho) & = 3 U_\ell''(\rho ) +12 \rho  U_\ell'''(\rho ) +4 \rho^2 U_\ell''''(\rho ) \nonumber \\
& \ \ \ {} + \left(k^2+p^2\right) \left[ Z_\ell'(\rho ) +2 \rho Z_\ell''(\rho) + 2 Y_\ell(\rho ) + 10 \rho Y_\ell'(\rho ) + 4 \rho ^2 Y_\ell''(\rho )\right] , \\
 \Gamma_\ell^{L,L,T,T} (-\bm{p},\bm{p},-\bm{k},\bm{k};\rho) & = U_\ell''(\rho ) +2 \rho  U_\ell'''(\rho ) + \left(k^2+p^2\right) Z_\ell'(\rho ) + 2 k^2 \rho Z_\ell''(\rho) + p^2 \rho Y_\ell'(\rho ) , \\
 \Gamma_\ell^{T,T,T',T'} (-\bm{p},\bm{p},-\bm{k},\bm{k};\rho) & = U_\ell''(\rho )  + \left(k^2+p^2\right) \left[Z_\ell'(\rho ) + 2 \delta_{T,T'} Y_\ell(\rho )\right] , \\ 
\Gamma_\ell^{L,L,L} (\bm{k}_1,\bm{k}_2,\bm{k}_3;\rho) & = \sqrt{2\rho} \left[
 3 U_\ell''(\rho ) + 2 \rho  U_\ell^{(3)}(\rho) + \frac{1}{2} \left(k_1^2+k_2^2+k_3^2\right)
   \left(Z_\ell'(\rho )+2 Y_\ell(\rho ) +2 \rho  Y_\ell'(\rho ) \right) \right] , \\
\Gamma_\ell^{L,T,T} (\bm{k}_1,\bm{k}_2,\bm{k}_3;\rho) & = \sqrt{2\rho} \left[
 U_\ell''(\rho ) - \bm{k}_2 \cdot \bm{k}_3 Z_\ell'(\rho ) + k_1^2 Y_\ell(\rho ) \right] , \\
\Gamma_{\ell;\mu \nu}^{A,A,L,L} (-\bm{p},\bm{p},-\bm{k},\bm{k};\rho) & = 
-\left(p^2 \delta_{\mu \nu} - p_\mu p_\nu \right)
 \left(\frac{(Z_\ell^A)'(\rho ) + 2 \rho (Z_\ell^A)''(\rho )}{e_\ell^2}\right)  - 2 \left[Z_\ell (\rho) + 5 \rho Z_\ell ' (\rho) + 2 \rho^2 Z_\ell '' (\rho) \right] \delta_{\mu \nu} , \\
\Gamma_{\ell;\mu \nu}^{A,A,T,T} (-\bm{p},\bm{p},-\bm{k},\bm{k};\rho) & = 
-\left(p^2 \delta_{\mu \nu} - p_\mu p_\nu \right)
 \left(\frac{(Z_\ell^A)'(\rho )}{e_\ell^2}\right)  - 2 \left[Z_\ell (\rho) + \rho Z_\ell ' (\rho) \right] \delta_{\mu \nu} , \\
 \Gamma_{\ell;\mu}^{A,L,T}  (\bm{k}_1,\bm{k}_2,\bm{k}_3;\rho) & =
i \Big( \left[k_{2,\mu} - k_{3,\mu} \right] Z_\ell (\rho) - 2 k_{3,\mu} \rho Z_\ell' (\rho) \Big) , \\  
 \Gamma_{\ell;\mu}^{A,T,T}  (\bm{k}_1,\bm{k}_2,\bm{k}_3;\rho) & =
i \left[k_{2,\mu} - k_{3,\mu} \right] Z_\ell (\rho)  , \\  
 \Gamma_{\ell;\mu}^{A,A,L}  (\bm{k}_1,\bm{k}_2,\bm{k}_3;\rho) & =
\sqrt{2\rho} \left( \left(\bm{k}_1 \cdot \bm{k}_2 \delta_{\mu \nu} - k_{1,\mu} k_{2,\nu} \right)  \left(\frac{(Z_\ell^A)'(\rho )}{e_\ell^2}\right) 
 - 2 \left[Z_\ell (\rho) + \rho Z_\ell ' (\rho) \right] \delta_{\mu \nu} \right) .
\end{align}
\end{widetext}
These vertices underly the constraint that the sum of all momenta should vanish. It should be noted that for the vertex $ \Gamma_\ell^{T,T,T',T'}$ it makes a difference whether the two transverse channels $T$ and $T'$ are identical or not.

\section{Rescaled flow equations}
\label{sec:RescaledFlowEq}

To derive dimensionless and rescaled flow equations, we use the scaling transformations given in Eqs.~(\ref{eq:scaling:x})--(\ref{eq:flowe2}). In addition, we introduce the rescaled momenta
\begin{equation}
	 \bm{q} = \bm{k}/\Lambda .
\end{equation}
For later reference, let us also define $\eta_{i,\ell} = \eta_\ell$ with $i=L,T$.
As concerns the regulators, it is convenient to write them as
\begin{align}
  \label{eq:dimlessregulator}
  R_\ell^\phi (k^2) &= Z_{\ell}^0 k^2 r_{\phi} (q^2)  , \\
  R_\ell^A (k^2) &= (1/e_\ell^2) k^2 r_{A} (q^2)  ,
\end{align}
where the $r_i(q^2)$ are dimensionless cutoff functions which do not explicitly depend on the cutoff $\Lambda$ (or the flow parameter $\ell$).
A convenient choice, which we will employ later on, is the Litim regulator \cite{Litim01}
\begin{equation}
  \label{eq:Litimcutoff}
  r_i(x) = \left(\frac{1}{x} -1 \right) \theta (1 - x) .
\end{equation}

In terms of the above dimensionless and rescaled quantities, the flow equation for the effective potential [see Eq.~(\ref{eq:Ubetafunction})] turns into \cite{Reuter94b,Bergerhoff96a,Bergerhoff96b}
\begin{align}
  \label{eq:flow:u}
  & \partial_\ell u (\tilde\rho) = D u  (\tilde\rho) - (D - 2 + \eta) \tilde \rho u'  (\tilde\rho) 
- ({K_D}/{2}) 
  \big[ L_{0,L}^D  (\tilde\rho) \nonumber \\
 & \qquad \quad \ \ {} + (2N - 1) L_{0,T}^D  (\tilde\rho) + (D-1) L_{0,A}^D  (\tilde\rho) \big] ,
\end{align}
where $K_D = \Omega_D/(2\pi)^D = 1/(2^{D-1} \pi^{D/2} \Gamma(D/2))$ is the surface area of a $D$-dimensional unit sphere divided by $(2\pi)^D$.
The threshold functions $L_{0,i}^D (\tilde\rho)$ occurring here are defined by
\begin{align}
  \label{eq:thresholdL0}
  L_{0,i}^D  (\tilde\rho) & = - \frac{1}{2} \int_0^{\infty} dx\, x^{D/2-1} \tilde G_i (x;\tilde\rho) \tilde \partial_\ell P_i (x;\tilde\rho) \nonumber \\
 & = - \frac{1}{2} \int_0^{\infty} dx\, x^{D/2-1} \tilde G_i (x;\tilde\rho) (\eta_i r_i (x) x + 2 r_i' (x) x^2) ,
\end{align}
where
\begin{equation}
  \label{eq:Gtilde}
  \tilde G_i (x;\tilde\rho) = \frac{1}{P_i (x;\tilde\rho) + w_i(\tilde\rho)} = \frac{1}{z_i (\tilde\rho) x + r_i(x) x + w_i(\tilde\rho) } 
\end{equation}
with $P_i (x;\tilde\rho) = z_i (\tilde\rho) x + r_i(x) x $ 
are the flowing rescaled propagators
and we have used
\begin{equation}
  \label{eq:derivativeofinverseprop}
  \tilde \partial_\ell P_i (x;\tilde\rho) = \eta_i r_i (x) x + 2 r_i' (x) x^2 .
\end{equation}
We have also defined
$w_T(\tilde\rho) = u'(\tilde\rho)$,
$w_L(\tilde\rho) = u'(\tilde\rho) + 2 \tilde\rho u''(\tilde\rho)$,
$w_A(\tilde\rho) = 2 \tilde e^2 \tilde\rho z (\tilde\rho)$,
$z_L(\tilde\rho) = \tilde z(\tilde\rho)$, and
$z_T(\tilde\rho) = z(\tilde\rho)$.

From a numerical standpoint, it is more convenient to consider the differential equation for $w(\tilde \rho) = w_T(\tilde \rho) = u'(\tilde \rho)$,
\begin{align}
  \label{eq:flow:w}
  & \partial_\ell w (\tilde\rho) = (2 - \eta) w (\tilde\rho) - (D - 2 + \eta) \tilde \rho w'  (\tilde\rho) \nonumber \\
  & \quad {} + ({K_D}/{2}) \Big[ w_L' L_{1,L}^D (\tilde\rho) +  z_L' L_{1,L}^{D+2} (\tilde\rho) \nonumber \\
 & \qquad \qquad  {} + (2N - 1)\left(  w_T' L_{1,T}^D (\tilde\rho) +  z_T' L_{1,T}^{D+2} (\tilde\rho) \right) \nonumber \\
 & \qquad \qquad  {} + (D-1) \left(  w_A' L_{1,A}^D (\tilde\rho) +  z_A' L_{1,A}^{D+2} (\tilde\rho) \right) \Big] .
\end{align}
After a straightforward but tedious calculation we obtain the dimensionless and rescaled flow equations for the wave-function renormalization factors $z (\tilde\rho) = z_\ell(\tilde\rho)$, $\tilde z (\tilde\rho) = \tilde z_\ell(\tilde\rho)$, and $z_A (\tilde\rho) = z_{\ell}^{A}(\tilde\rho)$, as well as a flow equation for the square of the dimensionless charge $\tilde e^2 = \tilde e_\ell^2$,
\begin{widetext}
 \begin{align}
 & \partial_\ell \tilde z (\tilde \rho)  = - \eta \tilde z (\tilde \rho) - (D -2 + \eta) \tilde \rho \tilde z' (\tilde \rho)
  + (K_D/2) \left[\tilde z'(\tilde \rho) + 2\tilde\rho  \tilde z''(\tilde \rho) \right] L_{1,L}^D(\tilde \rho)  \nonumber \\
  & \quad {} - 2 K_D \tilde\rho  \tilde z'(\tilde \rho)
 \left[3 u''(\tilde \rho) + 2 \tilde\rho  u'''(\tilde \rho)\right] L_{2,L,L}^D(\tilde \rho) 
 - (2 + 1/D) K_D \tilde\rho  \left[\tilde z'(\tilde \rho) \right]^2 L_{2,L,L}^{D+2}(\tilde \rho) \nonumber \\
 & \quad {} +(2/D) K_D \tilde\rho  \left[3 u''(\tilde \rho)+2 \tilde\rho  u'''(\tilde \rho)\right]^2 M_{2,L,L}^D(\tilde \rho)
 + (4/D) K_D \tilde\rho  \tilde z'(\tilde \rho) \left[3 u''(\tilde \rho)+2 \tilde\rho  u'''(\tilde \rho)\right] M_{2,L,L}^{D+2}(\tilde \rho) \nonumber \\
 & \quad {} +(2/D) K_D \tilde\rho  \left[\tilde z'(\tilde \rho) \right]^2 M_{2,L,L}^{D+4}(\tilde \rho) \nonumber \\ 
 & \quad {} - (2N-1) (K_D/2) \left[\left(\tilde z(\tilde \rho) - z(\tilde \rho)\right)/\tilde \rho - \tilde z'(\tilde \rho)\right] L_{1,T}^D(\tilde \rho) 
 - (2N-1) K_D \left[\tilde z (\tilde \rho) - z (\tilde \rho) \right] u''(\tilde \rho) L_{2,T,T}^D(\tilde \rho) \nonumber \\
 & \quad {} - (2N-1) K_D z'(\tilde \rho) \left[\tilde z(\tilde \rho) - z(\tilde \rho) +\tilde\rho  z'(\tilde \rho)/D\right] L_{2,T,T}^{D+2}(\tilde \rho)
 + (2/D) (2N-1) K_D \tilde\rho  \left[u''(\tilde \rho)\right]^2 M_{2,T,T}^D(\tilde \rho) \nonumber \\
 & \quad {} +(4/D) (2N-1)  K_D \tilde\rho z'(\tilde \rho) u''(\tilde \rho) M_{2,T,T}^{D+2}(\tilde \rho)
 + (2/D) (2N-1) K_D \tilde\rho  \left[z'(\tilde \rho)\right]^2 M_{2,T,T}^{D+4}(\tilde \rho) \nonumber \\
 & \quad {} +4 \left(1-{1}/{D}\right) K_D \tilde e^4 \tilde\rho  \left[z(\tilde \rho) + \tilde\rho  z'(\tilde \rho)\right]^2 L_{2,A,A}^{D-2}(\tilde \rho) 
 - 4 \left(1-{1}/{D}\right) K_D \tilde e^2 \left[ z(\tilde \rho) + \tilde\rho  z'(\tilde \rho)\right]^2 L_{2,A,T}^D(\tilde \rho) \nonumber \\ 
 & \quad {} - 2 \left(1- {1}/{D}\right) K_D \tilde\rho \left[ z_A'(\tilde \rho) \right]^2 L_{2,A,A}^{D+2}(\tilde \rho) 
 + 8 \left(1-{1}/{D}\right) K_D \tilde e^4 \tilde\rho  \left[z(\tilde \rho) + \tilde\rho  z'(\tilde \rho)\right]^2 M_{2,A,A}^D(\tilde \rho) \nonumber \\
  & \quad {} +8 \left(1-{1}/{D}\right) K_D \tilde e^2 \tilde\rho  \left[z(\tilde \rho) + \tilde\rho  z'(\tilde \rho)\right] z_A'(\tilde \rho) M_{2,A,A}^{D+2}(\tilde \rho) 
 + 2 \left(1-{1}/{D}\right) K_D \tilde\rho \left[z_A'(\tilde \rho)\right]^2 M_{2,A,A}^{D+4}(\tilde \rho) ,
 \end{align}

 \begin{align}
 & \quad \partial_\ell z (\tilde \rho)  = - \eta z (\tilde \rho) - (D -2 + \eta) \tilde \rho z' (\tilde \rho)
  + (K_D/2) \left[ z'(\tilde \rho) + 2 \tilde\rho  z''(\tilde \rho) \right] L_{1,L}^D(\tilde \rho) \nonumber \\
 & \quad {} +(K_D/2) \left[(\tilde z(\tilde \rho) -z(\tilde \rho))/\tilde \rho + (2N - 1)  z'(\tilde \rho)\right] L_{1,T}^D(\tilde \rho)   \nonumber \\
 & \quad {} - K_D \left[\tilde z(\tilde \rho)-z(\tilde \rho)+ 2 \tilde\rho  z'(\tilde \rho)\right] u''(\tilde \rho) L_{2,L,T}^D(\tilde \rho) \nonumber \\
 & \quad {} - K_D \left( (1/2+1/D) \left[\tilde z(\tilde \rho)-z(\tilde \rho)\right]^2/\tilde \rho +  (1-2/D) \left[\tilde z(\tilde \rho)- z(\tilde \rho)\right] z'(\tilde \rho) + (2/D) \tilde\rho \left[z'(\tilde \rho)\right]^2\right) L_{2,L,T}^{D+2}(\tilde \rho) \nonumber \\
 & \quad {} +(4/D) K_D \tilde\rho  \left[u''(\tilde \rho)\right]^2 M_{2,L,T}^D(\tilde \rho)
 + (4/D) K_D \left[\tilde z(\tilde \rho)- z(\tilde \rho)\right] u''(\tilde \rho) M_{2,L,T}^{D+2}(\tilde \rho)   \nonumber \\
 & \quad {} +(1/D) K_D \left( \left[\tilde z(\tilde \rho)- z(\tilde \rho) \right]^2/\tilde \rho \right) M_{2,L,T}^{D+4}(\tilde \rho)  \nonumber \\
 & \quad {} - (2/D) K_D  \left[\tilde z(\tilde \rho)  - z(\tilde \rho) - 2 \tilde\rho  z'(\tilde \rho)\right] u''(\tilde \rho) \left[ N_{2,L,T}^D(\tilde \rho) - N_{2,T,L}^D(\tilde \rho) \right] \nonumber \\
 & \quad {} - (1/D) K_D \left[ \tilde z(\tilde \rho) - z(\tilde \rho) \right] \left[\left(\tilde z(\tilde \rho) - z(\tilde \rho) \right)/\tilde \rho - 2 z'(\tilde \rho)\right] \left[ N_{2,L,T}^{D+2}(\tilde \rho) - N_{2,T,L}^{D+2}(\tilde \rho) \right] \nonumber \\
 & \quad {} - 4 K_D \left(1-{1}/{D}\right) \tilde e^2 \left[ z(\tilde \rho) + \tilde\rho  z'(\tilde \rho) \right]^2 L_{2,A,L}^D(\tilde \rho)  ,
 \end{align}

\begin{align}
& \partial_\ell z_A (\tilde\rho)  = - \eta_A z_A (\tilde\rho) - (D -2 + \eta) \tilde \rho z_A' (\tilde\rho)
+ (K_D/2) \left[ z_A'(\tilde\rho) + 2 \tilde\rho  z_A''(\tilde\rho )\right]  L_{1,L}^D(\tilde\rho ) \nonumber \\
& \quad {} +(2N-1) (K_D/2) z_A'(\tilde\rho )  L_{1,T}^D(\tilde\rho ) \nonumber \\
& \quad {} +\frac{8 K_D \tilde e^2}{D (D+2)} \left( \left[z(\tilde\rho ) + \tilde\rho  z'(\tilde\rho ) - \alpha_\phi \right]^2 M_{2,L,T}^{D+2}(\tilde\rho ) 
- 2 \alpha_\phi \eta_\phi \left[z(\tilde\rho ) + \tilde\rho  z'(\tilde\rho ) - \alpha_\phi \right] m_{2,L,T}^{D+2}(\tilde\rho ) \right)
\nonumber \\
& \quad {} +\frac{8  (N-1) K_D \tilde e^2}{D (D+2)} \left( \left[z(\tilde\rho) - \alpha_\phi \right]^2 M_{2,T,T}^{D+2}(\tilde\rho )  - 2 \alpha_\phi \eta_\phi \left[z(\tilde\rho ) - \alpha_\phi \right] m_{2,T,T}^{D+2}(\tilde\rho ) \right)\nonumber \\ 
& \quad {} - 4 K_D \tilde e^2\tilde\rho  \left[z(\tilde\rho ) + \tilde\rho  z'(\tilde\rho )\right] z_A'(\tilde\rho ) L_{2,A,L}^D(\tilde\rho ) 
- (4/D) K_D \tilde\rho \left[ z_A'(\tilde\rho ) \right]^2 L_{2,A,L}^{D+2}(\tilde\rho ) \nonumber \\
& \quad {} +\frac{16  (D+1) K_D \tilde e^4}{D(D+2)} \tilde\rho  \left[z(\tilde\rho ) + \tilde\rho  z'(\tilde\rho )\right]^2 M_{2,A,L}^D(\tilde\rho )  
%
- \frac{16 K_D \tilde e^4}{D(D+2)} \tilde\rho  \left[z(\tilde\rho ) + \tilde\rho  z'(\tilde\rho ) \right]^2 N_{2,A,L}^{D-2}(\tilde\rho ) \nonumber \\
& \quad {} - (8/D) K_D \tilde e^2 \tilde\rho  \left[z(\tilde\rho ) + \tilde\rho  z'(\tilde\rho )\right] z_A'(\tilde\rho ) \left[ N_{2,A,L}^D(\tilde\rho ) - N_{2,L,A}^D(\tilde\rho ) \right] \nonumber \\
%
%
%
& \quad  + (D-2) (K_D/6) \tilde e^2 \left[L_{\mathcal{C},L}^{D-2} (\tilde\rho) + (N-1) L_{\mathcal{C},T}^{D-2} (\tilde\rho) \right] ,
  \label{eq:flowzA}
\end{align}
\begin{equation}
  \label{eq:floweqdimcharge}
  \partial_\ell \tilde e^2 = (4 - D - \eta_A) \tilde e^2 .
\end{equation}
From the last equation, it follows directly that at a fixed point we must have \cite{Halperin74} $\eta_A = 4 - D$.
For an arbitrary regulator, the new threshold functions occurring above are given by
\begin{equation}
  \label{eq:thresholdL1}
  L_{1,i}^D  (\tilde\rho) = \frac{1}{2} \int_0^{\infty} dx\, x^{D/2-1} \tilde \partial_\ell \tilde G_i (x;\tilde\rho) = - \frac{1}{2} \int_0^{\infty} dx\, x^{D/2-1} \tilde G_i^2 (x;\tilde\rho)  (\eta_i r_i x + 2 r_i' x^2) ,
\end{equation}
\begin{align}
  \label{eq:thresholdL2}
  L_{2,i,j}^D  (\tilde\rho) &= \frac{1}{2} \int_0^{\infty} dx\, x^{D/2-1} \tilde \partial_\ell \left[ \tilde G_i (x;\tilde\rho) \tilde G_j (x;\tilde\rho) \right]  \nonumber \\
 &= - \frac{1}{2} \int_0^{\infty} dx\, x^{D/2-1} \tilde G_i (x;\tilde\rho) \tilde G_j (x;\tilde\rho) \left( \tilde G_i (x;\tilde\rho) (\eta_i r_i x + 2 r_i' x^2) + \tilde G_j (x;\tilde\rho)  (\eta_j r_j x + 2 r_j' x^2) \right) ,
\end{align}
\begin{align}
  \label{eq:thresholdM2}
 M_{2,i,j}^D  (\tilde\rho) &= \frac{1}{2} \int_0^{\infty} dx\, x^{D/2} \tilde \partial_\ell \left[ \tilde G_i ' (x;\tilde\rho) \tilde G_j ' (x;\tilde\rho) \right]  \nonumber \\
 & =  \frac{1}{2} \int_0^{\infty} dx\, x^{D/2}\, \tilde G_i^2 (x;\tilde\rho) \tilde G_j^2 (x;\tilde\rho) \Big[ 
\left(\eta_i (r_i +r_i' x) + 4 r_i' x + 2 r_i'' x^2 \right)  (z_j (\tilde\rho) + r_j + r_j' x) 
\nonumber \\
 & \qquad \qquad 
+ (z_i +r_i +r_i' x) \left(\eta_j (r_j +r_j' x) + 4 r_j' x + 2 r_j'' x^2 \right)  
\nonumber \\
&\qquad \qquad 
- 2 (z_i +r_i +r_i' x) (z_j (\tilde\rho) + r_j + r_j' x) \left(\tilde G_i (x;\tilde\rho) (\eta_i r_i x + 2 r_i' x^2) + \tilde G_j (x;\tilde\rho) (\eta_j r_j x + 2 r_j' x^2)  \right)      \Big] ,
\end{align}
\begin{align}
  \label{eq:thresholdN2}
  N_{2,i,j}^D  (\tilde\rho) &= -\frac{1}{2} \int_0^{\infty} dx\, x^{D/2} \tilde \partial_\ell \left[ \tilde G_i (x;\tilde\rho) \tilde G_j ' (x;\tilde\rho) \right]  \nonumber \\
 &= \frac{1}{2} \int_0^{\infty} dx\, x^{D/2}\, \Big[ \tilde G_i (x;\tilde\rho) \tilde G_j^2 (x;\tilde\rho) \left(\eta_j (r_j + r_j' x) + 4 r_j' x + 2 r_j'' x^2 \right)   \nonumber \\
& \qquad  \qquad 
- (z_j (\tilde\rho) +r_j +r_j' x) \left( \tilde G_i^2 (x;\tilde\rho) \tilde G_j^2 (x;\tilde\rho)  (\eta_i r_i x +2 r_i' x^2 )      
+ 2 \tilde G_i (x;\tilde\rho) \tilde G_j^3 (x;\tilde\rho) (\eta_j r_j x +2 r_j' x^2 ) \right)  
\Big]  ,
\end{align}
%
\begin{align}
  \label{eq:thresholdm2}
 m_{2,i,j}^D  (\tilde\rho) &= \frac{1}{2} \int_0^{\infty} dx\, x^{D/2} \left[ \tilde G_i ' (x;\tilde\rho) \tilde G_j ' (x;\tilde\rho) \right]  
=  \frac{1}{2} \int_0^{\infty} dx\, x^{D/2}\, \tilde G_i^2 (x;\tilde\rho) \tilde G_j^2 (x;\tilde\rho) (z_i +r_i +r_i' x) (z_j (\tilde\rho) + r_j + r_j' x) .
\end{align}
The last term in Eq.~(\ref{eq:flowzA}) is due to the correction term $\mathcal{C}_\ell [\phi,\bm{A}]$ and contains the threshold functions 
\begin{equation}
  \label{eq:thresholdLC}
  L_{\mathcal{C},i}^D (\tilde\rho) = \left[ 2 z_i(\tilde\rho) + 2 w_i(\tilde\rho) - \partial_\ell z_i (\tilde\rho) - \partial_\ell w_i (\tilde\rho) \right]  L_{\mathcal{C},i}^{D,a} (\tilde\rho) + 2 L_{\mathcal{C},i}^{D,b} (\tilde\rho) - L_{\mathcal{C},i}^{D,c} (\tilde\rho) ,
\end{equation}
with
\begin{align}
  \label{eq:thresholdLCa}
  & L_{\mathcal{C},i}^{D,a}  (\tilde\rho) = - \frac{1}{2} \int_0^{\infty} dx\, x^{D/2-1} \frac{r_i + r_i' x}{\left[ z_i(\tilde\rho) + w_i(\tilde\rho) + r_i x \right]^2} , \\
  \label{eq:thresholdLCb}
  & L_{\mathcal{C},i}^{D,b}  (\tilde\rho) = \frac{1}{2} \int_0^{\infty} dx\, x^{D/2-1} \frac{(r_i + r_i' x) r_i' x^2}{\left[ z_i(\tilde\rho) + w_i(\tilde\rho) + r_i x \right]^2} , \\ 
  \label{eq:thresholdLCc}
  & L_{\mathcal{C},i}^{D,c}  (\tilde\rho) = \frac{1}{2} \int_0^{\infty} dx\, x^{D/2-1} 
    \frac{4 r_i' x + 2 r_i'' x^2}{z_i(\tilde\rho) + w_i(\tilde\rho) + r_i x} .
\end{align}
\end{widetext}
Finally, to close our set of flow equations, we determine the flowing anomalous dimension $\eta_\ell$ of the field $\phi$ and its gauge-field counterpart $\eta_{\ell}^{A}$ by demanding that the corresponding wave-function renormalization factors evaluated at a characteristic value of the rescaled density $\tilde \rho = \tilde \rho_0$ are equal to one,
\begin{equation}
  \label{eq:etas}
  z_\ell (\tilde \rho_c^\ast) = z_{\ell}^{A} (\tilde \rho_c^\ast) = 1 .  
\end{equation}
While the definition of $\tilde \rho_c^\ast$ does not really matter (but effectively modifies the cutoff function), we find it convenient to choose it to be equal to the position of the critical rescaled effective potential $\tilde u^{\ast} (\tilde \rho)$, as shown in \Fig{fig:fixedpointfunctions}.

Using the Litim cutoff, as given in \Eq{eq:Litimcutoff}, its second derivative $r_i^{\prime \prime} (x)$ contains the derivative of a Dirac delta distribution which can be eliminated by an integration by parts. We then obtain for the threshold functions
\begin{widetext}
\begin{equation}
  \label{eq:thresholdL0Litim}
  L_{0,i}^D (\tilde\rho)  = \frac{1}{2} \int_0^{1} dx\, x^{D/2-1} \tilde G_i (x;\tilde\rho)  \alphaicancel (2- \eta_i (1-x)) ,
\end{equation}
\begin{equation}
  \label{eq:thresholdL1Litim}
  L_{1,i}^D (\tilde\rho)  = \frac{1}{2} \int_0^{1} dx\, x^{D/2-1} \tilde G_i^2 (x;\tilde\rho)  \alphaicancel (2- \eta_i (1-x)) ,
\end{equation}
\begin{align}
  \label{eq:thresholdL2Litim}
  & L_{2,i,j}^{D} (\tilde\rho) = \frac{1}{2} \int_0^1 dx\, x^{D/2-1} \tilde G_i (x;\tilde\rho) \tilde G_j (x;\tilde\rho) 
\left[ \tilde G_i (x;\tilde\rho) \alphaicancel ( 2- \eta_i(1-x) ) + \tilde G_j (x;\tilde\rho) \alphajcancel  ( 2- \eta_j(1-x) ) \right],
\end{align}
\begin{align}
  \label{eq:thresholdM2Litim}
  & M_{2,i,j}^{D} (\tilde\rho) = 
\left[z_i (\tilde\rho) z_j (\tilde\rho) -(z_i (\tilde\rho) - \alphaione)(z_j (\tilde\rho) - \alphajone) \right] 
\tilde G_i^2 (x=1;\tilde\rho) \tilde G_j^2 (x=1;\tilde\rho) 
+ \frac{1}{2} \int_0^1 dx\, x^{D/2} \tilde G_i^2 (x;\tilde\rho) \tilde G_j^2 (x;\tilde\rho) 
\nonumber \\
& \quad   {}\times \left[
- \alphaicancel \eta_i (z_j (\tilde\rho) - \alphajone) - \alphajcancel \eta_j (z_i (\tilde\rho) - \alphaione) +  2 (z_i (\tilde\rho) - \alphaione) (z_j (\tilde\rho) - \alphajone) \left(\tilde G_i (x;\tilde\rho) \alphaicancel (2 - \eta_i(1-x)) + \tilde G_j (x;\tilde\rho) \alphajcancel (2 - \eta_j(1-x)) \right) \right] ,
\end{align}
\begin{align}
  \label{eq:thresholdN2Litim}
  & N_{2,i,j}^{D} (\tilde\rho) = \alphajcancel \tilde G_i (x=1;\tilde\rho) \tilde G_j^2 (x=1;\tilde\rho) 
+ \frac{1}{2} \int_0^1 dx\, x^{D/2} \Big[
-  \tilde G_i (x;\tilde\rho) \tilde G_j^2 (x;\tilde\rho) \alphajcancel \eta_j  \nonumber \\
& \qquad \qquad \qquad  {} 
 + (z_j (\tilde\rho) - \alphajone) \left( \tilde G_i^2 (x;\tilde\rho) \tilde G_j^2 (x;\tilde\rho) \alphaicancel (2 - \eta_i(1-x))
 + 2 \tilde G_i (x;\tilde\rho) \tilde G_j^3 (x;\tilde\rho) \alphajcancel  (2 - \eta_j(1-x)) \right) \Big] ,
\end{align}
\begin{align}
  \label{eq:thresholdm2Litim}
  & m_{2,i,j}^{D} (\tilde\rho) = 
+ (z_i (\tilde\rho) - \alphaione) (z_j (\tilde\rho) - \alphajone) \frac{1}{2} \int_0^1 dx\, x^{D/2} \tilde G_i^2 (x;\tilde\rho) \tilde G_j^2 (x;\tilde\rho) ,
\end{align}
\begin{align}
  \label{eq:thresholdLCaLitim}
  L_{\mathcal{C},i}^{D,a}  (\tilde\rho) & =   L_{\mathcal{C},i}^{D,b}  (\tilde\rho) 
  = \frac{1}{2} \int_0^{1} dx\, x^{D/2-1} \frac{\alphaione}{\left[ z_i(\tilde\rho) + w_i(\tilde\rho) + \alphaicancel(1-x) \right]^2} , \\
  \label{eq:thresholdLCcLitim}
  L_{\mathcal{C},i}^{D,c}  (\tilde\rho) & = \frac{\alphaione}{z_i(\tilde\rho) + w_i(\tilde\rho)} .
%
\end{align}
Collecting all terms for $ L_{\mathcal{C},i}^{D,a}  (\tilde\rho)$, we obtain
\begin{equation}
  \label{eq:thresholdLCLitim}
  L_{\mathcal{C},i}^D (\tilde\rho) = \left[ 2 \left(\alphaione +  z_i(\tilde\rho) + w_i(\tilde\rho) \right) - \partial_\ell z_i (\tilde\rho) - \partial_\ell w_i \right (\tilde\rho) ]  L_{\mathcal{C},i}^{D,a} (\tilde\rho) - \frac{\alphaione}{z_i(\tilde\rho) + w_i(\tilde\rho)} .
\end{equation}
In the case of $D=3$, we only need $L_{\mathcal{C},i}^{D = 1,a} (\tilde\rho)$, which in fact is easily calculated analytically,
\begin{equation}
  \label{eq:thresholdLCaLitimD=1}
 L_{\mathcal{C},i}^{D = 1,a}  (\tilde\rho) = \frac{\alphaione}{2 [z_i(\tilde\rho) + w_i(\tilde\rho)] \left[\alphaione+z_i(\tilde\rho)+w_i(\tilde\rho)\right]} +\frac{
 1}{2 \left[\alphaione+z_i(\tilde\rho)+w_i(\tilde\rho)\right]^{3/2}} 
\text{Artanh} \left(\sqrt{\frac{\alphaione}{\alphaione+z_i(\tilde\rho)+w_i(\tilde\rho)} }\right) . \\
\end{equation}
\end{widetext}
To solve our complete set of flow equations, we discretize $\tilde\rho$, evaluate the threshold functions using quadrature and advance the solution using a fourth order Runge-Kutta method with a sufficiently small step size. 
For the results displayed in Figs.~\ref{fig:fixedpointfunctions}--\ref{fig:Evolution2} we have used 201 points equally spaced between $\tilde\rho = 0$ and $2$ ($N=1$) or between $\tilde\rho = 0$ and $2.83$ ($N=2$), respectively. The step size in the Runge-Kutta solver was $\delta \ell = 0.001$.  
Starting from $\tilde e_0^2 = 0.1$ and $\tilde \lambda_0 = 1$, it was then possible to fine-tune to criticality using the bisection method.


\begin{thebibliography}{46}
\expandafter\ifx\csname natexlab\endcsname\relax\def\natexlab#1{#1}\fi
\expandafter\ifx\csname bibnamefont\endcsname\relax
  \def\bibnamefont#1{#1}\fi
\expandafter\ifx\csname bibfnamefont\endcsname\relax
  \def\bibfnamefont#1{#1}\fi
\expandafter\ifx\csname citenamefont\endcsname\relax
  \def\citenamefont#1{#1}\fi
\expandafter\ifx\csname url\endcsname\relax
  \def\url#1{\texttt{#1}}\fi
\expandafter\ifx\csname urlprefix\endcsname\relax\def\urlprefix{URL }\fi
\providecommand{\bibinfo}[2]{#2}
\providecommand{\eprint}[2][]{\url{#2}}

\bibitem[{\citenamefont{Senthil
  et~al.}(2004{\natexlab{a}})\citenamefont{Senthil, Vishwanath, Balents,
  Sachdev, and Fisher}}]{Senthil04a}
\bibinfo{author}{\bibfnamefont{T.}~\bibnamefont{Senthil}},
  \bibinfo{author}{\bibfnamefont{A.}~\bibnamefont{Vishwanath}},
  \bibinfo{author}{\bibfnamefont{L.}~\bibnamefont{Balents}},
  \bibinfo{author}{\bibfnamefont{S.}~\bibnamefont{Sachdev}}, \bibnamefont{and}
  \bibinfo{author}{\bibfnamefont{M.~P.~A.} \bibnamefont{Fisher}},
  \bibinfo{journal}{Science} \textbf{\bibinfo{volume}{303}},
  \bibinfo{pages}{1490} (\bibinfo{year}{2004}{\natexlab{a}}).

\bibitem[{\citenamefont{Senthil
  et~al.}(2004{\natexlab{b}})\citenamefont{Senthil, Balents, Sachdev,
  Vishwanath, and Fisher}}]{Senthil04b}
\bibinfo{author}{\bibfnamefont{T.}~\bibnamefont{Senthil}},
  \bibinfo{author}{\bibfnamefont{L.}~\bibnamefont{Balents}},
  \bibinfo{author}{\bibfnamefont{S.}~\bibnamefont{Sachdev}},
  \bibinfo{author}{\bibfnamefont{A.}~\bibnamefont{Vishwanath}},
  \bibnamefont{and} \bibinfo{author}{\bibfnamefont{M.~P.~A.}
  \bibnamefont{Fisher}}, \bibinfo{journal}{Phys. Rev. B}
  \textbf{\bibinfo{volume}{70}}, \bibinfo{pages}{144407}
  (\bibinfo{year}{2004}{\natexlab{b}}).

\bibitem[{\citenamefont{Sachdev}(2011)}]{SachdevBook}
\bibinfo{author}{\bibfnamefont{S.}~\bibnamefont{Sachdev}},
  \emph{\bibinfo{title}{{Quantum Phase Transitions}}}
  (\bibinfo{publisher}{Cambridge University Press, Cambridge},
  \bibinfo{year}{2011}), \bibinfo{edition}{2nd} ed.

\bibitem[{\citenamefont{Read and Sachdev}(1989)}]{Read89}
\bibinfo{author}{\bibfnamefont{N.}~\bibnamefont{Read}} \bibnamefont{and}
  \bibinfo{author}{\bibfnamefont{S.}~\bibnamefont{Sachdev}},
  \bibinfo{journal}{Phys. Rev. Lett.} \textbf{\bibinfo{volume}{62}},
  \bibinfo{pages}{1694} (\bibinfo{year}{1989}).

\bibitem[{\citenamefont{Read and Sachdev}(1990)}]{Read90}
\bibinfo{author}{\bibfnamefont{N.}~\bibnamefont{Read}} \bibnamefont{and}
  \bibinfo{author}{\bibfnamefont{S.}~\bibnamefont{Sachdev}},
  \bibinfo{journal}{Phys. Rev. B} \textbf{\bibinfo{volume}{42}},
  \bibinfo{pages}{4568} (\bibinfo{year}{1990}).

\bibitem[{\citenamefont{Levin and Senthil}(2004)}]{Levin04}
\bibinfo{author}{\bibfnamefont{M.}~\bibnamefont{Levin}} \bibnamefont{and}
  \bibinfo{author}{\bibfnamefont{T.}~\bibnamefont{Senthil}},
  \bibinfo{journal}{Phys. Rev. B} \textbf{\bibinfo{volume}{70}},
  \bibinfo{pages}{220403} (\bibinfo{year}{2004}).

\bibitem[{\citenamefont{Te{\v{s}}anovi{\'c}}(2004)}]{Tesanovic04}
\bibinfo{author}{\bibfnamefont{Z.}~\bibnamefont{Te{\v{s}}anovi{\'c}}},
  \bibinfo{journal}{Phys. Rev. Lett.} \textbf{\bibinfo{volume}{93}},
  \bibinfo{pages}{217004} (\bibinfo{year}{2004}).

\bibitem[{\citenamefont{Senthil et~al.}(2005)\citenamefont{Senthil, Balents,
  Sachdev, Vishwanath, and Fisher}}]{Senthil05}
\bibinfo{author}{\bibfnamefont{T.}~\bibnamefont{Senthil}},
  \bibinfo{author}{\bibfnamefont{L.}~\bibnamefont{Balents}},
  \bibinfo{author}{\bibfnamefont{S.}~\bibnamefont{Sachdev}},
  \bibinfo{author}{\bibfnamefont{A.}~\bibnamefont{Vishwanath}},
  \bibnamefont{and} \bibinfo{author}{\bibfnamefont{M.~P.~A.}
  \bibnamefont{Fisher}}, \bibinfo{journal}{J. Phys. Soc. Jpn. Suppl.}
  \textbf{\bibinfo{volume}{74}}, \bibinfo{pages}{1} (\bibinfo{year}{2005}).

\bibitem[{\citenamefont{Balents
  et~al.}(2005{\natexlab{a}})\citenamefont{Balents, Bartosch, Burkov, Sachdev,
  and Sengupta}}]{Balents05a}
\bibinfo{author}{\bibfnamefont{L.}~\bibnamefont{Balents}},
  \bibinfo{author}{\bibfnamefont{L.}~\bibnamefont{Bartosch}},
  \bibinfo{author}{\bibfnamefont{A.}~\bibnamefont{Burkov}},
  \bibinfo{author}{\bibfnamefont{S.}~\bibnamefont{Sachdev}}, \bibnamefont{and}
  \bibinfo{author}{\bibfnamefont{K.}~\bibnamefont{Sengupta}},
  \bibinfo{journal}{Phys. Rev. B} \textbf{\bibinfo{volume}{71}},
  \bibinfo{pages}{144508} (\bibinfo{year}{2005}{\natexlab{a}}).

\bibitem[{\citenamefont{Balents
  et~al.}(2005{\natexlab{b}})\citenamefont{Balents, Bartosch, Burkov, Sachdev,
  and Sengupta}}]{Balents05b}
\bibinfo{author}{\bibfnamefont{L.}~\bibnamefont{Balents}},
  \bibinfo{author}{\bibfnamefont{L.}~\bibnamefont{Bartosch}},
  \bibinfo{author}{\bibfnamefont{A.}~\bibnamefont{Burkov}},
  \bibinfo{author}{\bibfnamefont{S.}~\bibnamefont{Sachdev}}, \bibnamefont{and}
  \bibinfo{author}{\bibfnamefont{K.}~\bibnamefont{Sengupta}},
  \bibinfo{journal}{Phys. Rev. B} \textbf{\bibinfo{volume}{71}},
  \bibinfo{pages}{144509} (\bibinfo{year}{2005}{\natexlab{b}}).

\bibitem[{\citenamefont{Balents
  et~al.}(2005{\natexlab{c}})\citenamefont{Balents, Bartosch, Burkov, Sachdev,
  and Sengupta}}]{Balents05c}
\bibinfo{author}{\bibfnamefont{L.}~\bibnamefont{Balents}},
  \bibinfo{author}{\bibfnamefont{L.}~\bibnamefont{Bartosch}},
  \bibinfo{author}{\bibfnamefont{A.}~\bibnamefont{Burkov}},
  \bibinfo{author}{\bibfnamefont{S.}~\bibnamefont{Sachdev}}, \bibnamefont{and}
  \bibinfo{author}{\bibfnamefont{K.}~\bibnamefont{Sengupta}},
  \bibinfo{journal}{Prog. Theor. Phys. Suppl.} \textbf{\bibinfo{volume}{160}},
  \bibinfo{pages}{314} (\bibinfo{year}{2005}{\natexlab{c}}).

\bibitem[{\citenamefont{Halperin et~al.}(1974)\citenamefont{Halperin, Lubensky,
  and Ma}}]{Halperin74}
\bibinfo{author}{\bibfnamefont{B.~I.} \bibnamefont{Halperin}},
  \bibinfo{author}{\bibfnamefont{T.~C.} \bibnamefont{Lubensky}},
  \bibnamefont{and} \bibinfo{author}{\bibfnamefont{S.-k.} \bibnamefont{Ma}},
  \bibinfo{journal}{Phys. Rev. Lett.} \textbf{\bibinfo{volume}{32}},
  \bibinfo{pages}{292} (\bibinfo{year}{1974}).

\bibitem[{\citenamefont{Irkhin et~al.}(1996)\citenamefont{Irkhin, Katanin, and
  Katsnelson}}]{Irkhin96}
\bibinfo{author}{\bibfnamefont{V.~Y.} \bibnamefont{Irkhin}},
  \bibinfo{author}{\bibfnamefont{A.~A.} \bibnamefont{Katanin}},
  \bibnamefont{and} \bibinfo{author}{\bibfnamefont{M.~I.}
  \bibnamefont{Katsnelson}}, \bibinfo{journal}{Phys. Rev. B}
  \textbf{\bibinfo{volume}{54}}, \bibinfo{pages}{11953} (\bibinfo{year}{1996}).

\bibitem[{\citenamefont{Kaul and Sachdev}(2008)}]{Kaul08}
\bibinfo{author}{\bibfnamefont{R.~K.} \bibnamefont{Kaul}} \bibnamefont{and}
  \bibinfo{author}{\bibfnamefont{S.}~\bibnamefont{Sachdev}},
  \bibinfo{journal}{Phys. Rev. B} \textbf{\bibinfo{volume}{77}},
  \bibinfo{pages}{155105} (\bibinfo{year}{2008}).

\bibitem[{\citenamefont{Dasgupta and Halperin}(1981)}]{Dasgupta81}
\bibinfo{author}{\bibfnamefont{C.}~\bibnamefont{Dasgupta}} \bibnamefont{and}
  \bibinfo{author}{\bibfnamefont{B.~I.} \bibnamefont{Halperin}},
  \bibinfo{journal}{Phys. Rev. Lett.} \textbf{\bibinfo{volume}{47}},
  \bibinfo{pages}{1556} (\bibinfo{year}{1981}).

\bibitem[{\citenamefont{Kiometzis et~al.}(1994)\citenamefont{Kiometzis,
  Kleinert, and Schakel}}]{Kiometzis94}
\bibinfo{author}{\bibfnamefont{M.}~\bibnamefont{Kiometzis}},
  \bibinfo{author}{\bibfnamefont{H.}~\bibnamefont{Kleinert}}, \bibnamefont{and}
  \bibinfo{author}{\bibfnamefont{A.~M.~J.} \bibnamefont{Schakel}},
  \bibinfo{journal}{Phys. Rev. Lett.} \textbf{\bibinfo{volume}{73}},
  \bibinfo{pages}{1975} (\bibinfo{year}{1994}).

\bibitem[{\citenamefont{Radzihovsky}(1995)}]{Radzihovsky95}
\bibinfo{author}{\bibfnamefont{L.}~\bibnamefont{Radzihovsky}},
  \bibinfo{journal}{Europhys. Lett.} \textbf{\bibinfo{volume}{29}},
  \bibinfo{pages}{227} (\bibinfo{year}{1995}).

\bibitem[{\citenamefont{Bergerhoff
  et~al.}(1996{\natexlab{a}})\citenamefont{Bergerhoff, Freire, Litim, Lola, and
  Wetterich}}]{Bergerhoff96a}
\bibinfo{author}{\bibfnamefont{B.}~\bibnamefont{Bergerhoff}},
  \bibinfo{author}{\bibfnamefont{F.}~\bibnamefont{Freire}},
  \bibinfo{author}{\bibfnamefont{D.~F.} \bibnamefont{Litim}},
  \bibinfo{author}{\bibfnamefont{S.}~\bibnamefont{Lola}}, \bibnamefont{and}
  \bibinfo{author}{\bibfnamefont{C.}~\bibnamefont{Wetterich}},
  \bibinfo{journal}{Phys. Rev. B} \textbf{\bibinfo{volume}{53}},
  \bibinfo{pages}{5734} (\bibinfo{year}{1996}{\natexlab{a}}).

\bibitem[{\citenamefont{Bergerhoff
  et~al.}(1996{\natexlab{b}})\citenamefont{Bergerhoff, Litim, Lola, and
  Wetterich}}]{Bergerhoff96b}
\bibinfo{author}{\bibfnamefont{B.}~\bibnamefont{Bergerhoff}},
  \bibinfo{author}{\bibfnamefont{D.~F.} \bibnamefont{Litim}},
  \bibinfo{author}{\bibfnamefont{S.}~\bibnamefont{Lola}}, \bibnamefont{and}
  \bibinfo{author}{\bibfnamefont{C.}~\bibnamefont{Wetterich}},
  \bibinfo{journal}{Int. J. Mod. Phys. A} \textbf{\bibinfo{volume}{11}},
  \bibinfo{pages}{4273} (\bibinfo{year}{1996}{\natexlab{b}}).

\bibitem[{\citenamefont{Folk and Holovatch}(1996)}]{Folk96}
\bibinfo{author}{\bibfnamefont{R.}~\bibnamefont{Folk}} \bibnamefont{and}
  \bibinfo{author}{\bibfnamefont{Y.}~\bibnamefont{Holovatch}},
  \bibinfo{journal}{J. Phys. A: Math. Gen.} \textbf{\bibinfo{volume}{29}},
  \bibinfo{pages}{3409} (\bibinfo{year}{1996}).

\bibitem[{\citenamefont{Herbut and Te{\v{s}}anovi{\'c}}(1996)}]{Herbut96}
\bibinfo{author}{\bibfnamefont{I.~F.} \bibnamefont{Herbut}} \bibnamefont{and}
  \bibinfo{author}{\bibfnamefont{Z.}~\bibnamefont{Te{\v{s}}anovi{\'c}}},
  \bibinfo{journal}{Phys. Rev. Lett.} \textbf{\bibinfo{volume}{76}},
  \bibinfo{pages}{4588} (\bibinfo{year}{1996}).

\bibitem[{\citenamefont{Kolnberger and Folk}(1990)}]{Kolnberger90}
\bibinfo{author}{\bibfnamefont{S.}~\bibnamefont{Kolnberger}} \bibnamefont{and}
  \bibinfo{author}{\bibfnamefont{R.}~\bibnamefont{Folk}},
  \bibinfo{journal}{Phys. Rev. B} \textbf{\bibinfo{volume}{41}},
  \bibinfo{pages}{4083} (\bibinfo{year}{1990}).

\bibitem[{\citenamefont{Lawrie and Athrone}(1983)}]{Lawrie83}
\bibinfo{author}{\bibfnamefont{I.~D.} \bibnamefont{Lawrie}} \bibnamefont{and}
  \bibinfo{author}{\bibfnamefont{C.}~\bibnamefont{Athrone}},
  \bibinfo{journal}{J. Phys. A} \textbf{\bibinfo{volume}{16}},
  \bibinfo{pages}{L587} (\bibinfo{year}{1983}).

\bibitem[{\citenamefont{Motrunich and Vishwanath}(2004)}]{Motrunich04}
\bibinfo{author}{\bibfnamefont{O.~I.} \bibnamefont{Motrunich}}
  \bibnamefont{and}
  \bibinfo{author}{\bibfnamefont{A.}~\bibnamefont{Vishwanath}},
  \bibinfo{journal}{Phys. Rev. B} \textbf{\bibinfo{volume}{70}},
  \bibinfo{pages}{075104} (\bibinfo{year}{2004}).

\bibitem[{\citenamefont{Sandvik}(2007)}]{Sandvik07}
\bibinfo{author}{\bibfnamefont{A.~W.} \bibnamefont{Sandvik}},
  \bibinfo{journal}{Phys. Rev. Lett.} \textbf{\bibinfo{volume}{98}},
  \bibinfo{pages}{227202} (\bibinfo{year}{2007}).

\bibitem[{\citenamefont{Nogueira et~al.}(2007)\citenamefont{Nogueira, Kragset,
  and Sudb\o{}}}]{Nogueira07}
\bibinfo{author}{\bibfnamefont{F.~S.} \bibnamefont{Nogueira}},
  \bibinfo{author}{\bibfnamefont{S.}~\bibnamefont{Kragset}}, \bibnamefont{and}
  \bibinfo{author}{\bibfnamefont{A.}~\bibnamefont{Sudb\o{}}},
  \bibinfo{journal}{Phys. Rev. B} \textbf{\bibinfo{volume}{76}},
  \bibinfo{pages}{220403} (\bibinfo{year}{2007}).

\bibitem[{\citenamefont{Melko and Kaul}(2008)}]{Melko08}
\bibinfo{author}{\bibfnamefont{R.~G.} \bibnamefont{Melko}} \bibnamefont{and}
  \bibinfo{author}{\bibfnamefont{R.~K.} \bibnamefont{Kaul}},
  \bibinfo{journal}{Phys. Rev. Lett.} \textbf{\bibinfo{volume}{100}},
  \bibinfo{pages}{017203} (\bibinfo{year}{2008}).

\bibitem[{\citenamefont{Motrunich and Vishwanath}(2008)}]{Motrunich08}
\bibinfo{author}{\bibfnamefont{O.~I.} \bibnamefont{Motrunich}}
  \bibnamefont{and}
  \bibinfo{author}{\bibfnamefont{A.}~\bibnamefont{Vishwanath}},
  \bibinfo{journal}{arXiv:0805.1494}  (\bibinfo{year}{2008}).

\bibitem[{\citenamefont{Sandvik}(2010)}]{Sandvik10}
\bibinfo{author}{\bibfnamefont{A.~W.} \bibnamefont{Sandvik}},
  \bibinfo{journal}{Phys. Rev. Lett.} \textbf{\bibinfo{volume}{104}},
  \bibinfo{pages}{177201} (\bibinfo{year}{2010}).

\bibitem[{\citenamefont{Kaul}(2011)}]{Kaul11}
\bibinfo{author}{\bibfnamefont{R.~K.} \bibnamefont{Kaul}},
  \bibinfo{journal}{Phys. Rev. B} \textbf{\bibinfo{volume}{84}},
  \bibinfo{pages}{054407} (\bibinfo{year}{2011}).

\bibitem[{\citenamefont{Pujari et~al.}(2013)\citenamefont{Pujari, Damle, and
  Alet}}]{Pujari13}
\bibinfo{author}{\bibfnamefont{S.}~\bibnamefont{Pujari}},
  \bibinfo{author}{\bibfnamefont{K.}~\bibnamefont{Damle}}, \bibnamefont{and}
  \bibinfo{author}{\bibfnamefont{F.}~\bibnamefont{Alet}},
  \bibinfo{journal}{Phys. Rev. Lett.} \textbf{\bibinfo{volume}{111}},
  \bibinfo{pages}{087203} (\bibinfo{year}{2013}).


\bibitem[{\citenamefont{Block et~al.}(2013)\citenamefont{Block, Melko, and
  Kaul}}]{Block13}
\bibinfo{author}{\bibfnamefont{M.~S.} \bibnamefont{Block}},
  \bibinfo{author}{\bibfnamefont{R.~G.} \bibnamefont{Melko}}, \bibnamefont{and}
  \bibinfo{author}{\bibfnamefont{R.~K.} \bibnamefont{Kaul}},
  \bibinfo{journal}{Phys. Rev. Lett.} \textbf{\bibinfo{volume}{111}},
  \bibinfo{pages}{137202} (\bibinfo{year}{2013}).

\bibitem[{\citenamefont{Harada et~al.}(2013)\citenamefont{Harada, Suzuki,
  Okubo, Matsuo, Lou, Watanabe, Todo, and Kawashima}}]{Harada13}
\bibinfo{author}{\bibfnamefont{K.}~\bibnamefont{Harada}},
  \bibinfo{author}{\bibfnamefont{T.}~\bibnamefont{Suzuki}},
  \bibinfo{author}{\bibfnamefont{T.}~\bibnamefont{Okubo}},
  \bibinfo{author}{\bibfnamefont{H.}~\bibnamefont{Matsuo}},
  \bibinfo{author}{\bibfnamefont{J.}~\bibnamefont{Lou}},
  \bibinfo{author}{\bibfnamefont{H.}~\bibnamefont{Watanabe}},
  \bibinfo{author}{\bibfnamefont{S.}~\bibnamefont{Todo}}, \bibnamefont{and}
  \bibinfo{author}{\bibfnamefont{N.}~\bibnamefont{Kawashima}},
  \bibinfo{journal}{arXiv:1307.0501}  (\bibinfo{year}{2013}).

\bibitem[{\citenamefont{Jiang et~al.}(2008)\citenamefont{Jiang, Nyfeler,
  Chandrasekharan, and Wiese}}]{Jiang08}
\bibinfo{author}{\bibfnamefont{F.-J.} \bibnamefont{Jiang}},
  \bibinfo{author}{\bibfnamefont{M.}~\bibnamefont{Nyfeler}},
  \bibinfo{author}{\bibfnamefont{S.}~\bibnamefont{Chandrasekharan}},
  \bibnamefont{and} \bibinfo{author}{\bibfnamefont{U.-J.} \bibnamefont{Wiese}},
  \bibinfo{journal}{J. Stat. Mech: Theory Exp.} p. \bibinfo{pages}{P02009}
  (\bibinfo{year}{2008}).

\bibitem[{\citenamefont{Kuklov et~al.}(2008)\citenamefont{Kuklov, Matsumoto,
  Prokof'ev, Svistunov, and Troyer}}]{Kuklov08}
\bibinfo{author}{\bibfnamefont{A.~B.} \bibnamefont{Kuklov}},
  \bibinfo{author}{\bibfnamefont{M.}~\bibnamefont{Matsumoto}},
  \bibinfo{author}{\bibfnamefont{N.~V.} \bibnamefont{Prokof'ev}},
  \bibinfo{author}{\bibfnamefont{B.~V.} \bibnamefont{Svistunov}},
  \bibnamefont{and} \bibinfo{author}{\bibfnamefont{M.}~\bibnamefont{Troyer}},
  \bibinfo{journal}{Phys. Rev. Lett.} \textbf{\bibinfo{volume}{101}},
  \bibinfo{pages}{050405} (\bibinfo{year}{2008}).

\bibitem[{\citenamefont{Herland et~al.}(2013)\citenamefont{Herland, Bojesen,
  Babaev, and Sudb\o{}}}]{Herland13}
\bibinfo{author}{\bibfnamefont{E.~V.} \bibnamefont{Herland}},
  \bibinfo{author}{\bibfnamefont{T.~A.} \bibnamefont{Bojesen}},
  \bibinfo{author}{\bibfnamefont{E.}~\bibnamefont{Babaev}}, \bibnamefont{and}
  \bibinfo{author}{\bibfnamefont{A.}~\bibnamefont{Sudb\o{}}},
  \bibinfo{journal}{Phys. Rev. B} \textbf{\bibinfo{volume}{87}},
  \bibinfo{pages}{134503} (\bibinfo{year}{2013}).

\bibitem[{\citenamefont{Chen et~al.}(2013)\citenamefont{Chen, Huang, Deng,
  Kuklov, Prokof'ev, and Svistunov}}]{Chen13}
\bibinfo{author}{\bibfnamefont{K.}~\bibnamefont{Chen}},
  \bibinfo{author}{\bibfnamefont{Y.}~\bibnamefont{Huang}},
  \bibinfo{author}{\bibfnamefont{Y.}~\bibnamefont{Deng}},
  \bibinfo{author}{\bibfnamefont{A.~B.} \bibnamefont{Kuklov}},
  \bibinfo{author}{\bibfnamefont{N.~V.} \bibnamefont{Prokof'ev}},
  \bibnamefont{and} \bibinfo{author}{\bibfnamefont{B.~V.}
  \bibnamefont{Svistunov}}, \bibinfo{journal}{Phys. Rev. Lett.}
  \textbf{\bibinfo{volume}{110}}, \bibinfo{pages}{185701}
  (\bibinfo{year}{2013}).

\bibitem[{\citenamefont{Berges et~al.}(2002)\citenamefont{Berges, Tetradis, and
  Wetterich}}]{Berges02}
\bibinfo{author}{\bibfnamefont{J.}~\bibnamefont{Berges}},
  \bibinfo{author}{\bibfnamefont{N.}~\bibnamefont{Tetradis}}, \bibnamefont{and}
  \bibinfo{author}{\bibfnamefont{C.}~\bibnamefont{Wetterich}},
  \bibinfo{journal}{Phys. Rep.} \textbf{\bibinfo{volume}{363}},
  \bibinfo{pages}{223} (\bibinfo{year}{2002}).

\bibitem[{\citenamefont{Kopietz et~al.}(2010)\citenamefont{Kopietz, Bartosch,
  and Sch\"{u}tz}}]{Kopietz10}
\bibinfo{author}{\bibfnamefont{P.}~\bibnamefont{Kopietz}},
  \bibinfo{author}{\bibfnamefont{L.}~\bibnamefont{Bartosch}}, \bibnamefont{and}
  \bibinfo{author}{\bibfnamefont{F.}~\bibnamefont{Sch\"{u}tz}},
  \emph{\bibinfo{title}{{Introduction to the Functional Renormalization
  Group}}} (\bibinfo{publisher}{Springer Verlag, Berlin},
  \bibinfo{year}{2010}).

\bibitem[{\citenamefont{Delamotte}(2012)}]{Delamotte12}
\bibinfo{author}{\bibfnamefont{B.}~\bibnamefont{Delamotte}}, in
  \emph{\bibinfo{booktitle}{Renormalization Group and Effective Field Theory
  Approaches to Many-Body Systems}}, edited by
  \bibinfo{editor}{\bibfnamefont{A.}~\bibnamefont{Schwenk}} \bibnamefont{and}
  \bibinfo{editor}{\bibfnamefont{J.}~\bibnamefont{Polonyi}}
  (\bibinfo{publisher}{Springer Berlin Heidelberg}, \bibinfo{year}{2012}), vol.
  \bibinfo{volume}{852} of \emph{\bibinfo{series}{Lecture Notes in Physics}},
  pp. \bibinfo{pages}{49--132}.

\bibitem[{\citenamefont{Gies}(2012)}]{Gies12}
\bibinfo{author}{\bibfnamefont{H.}~\bibnamefont{Gies}}, in
  \emph{\bibinfo{booktitle}{Renormalization Group and Effective Field Theory
  Approaches to Many-Body Systems}}, edited by
  \bibinfo{editor}{\bibfnamefont{A.}~\bibnamefont{Schwenk}} \bibnamefont{and}
  \bibinfo{editor}{\bibfnamefont{J.}~\bibnamefont{Polonyi}}
  (\bibinfo{publisher}{Springer Berlin Heidelberg}, \bibinfo{year}{2012}), vol.
  \bibinfo{volume}{852} of \emph{\bibinfo{series}{Lecture Notes in Physics}},
  pp. \bibinfo{pages}{287--348}.

\bibitem[{\citenamefont{Metzner et~al.}(2012)\citenamefont{Metzner, Salmhofer,
  Honerkamp, Meden, and Sch\"onhammer}}]{Metzner12}
\bibinfo{author}{\bibfnamefont{W.}~\bibnamefont{Metzner}},
  \bibinfo{author}{\bibfnamefont{M.}~\bibnamefont{Salmhofer}},
  \bibinfo{author}{\bibfnamefont{C.}~\bibnamefont{Honerkamp}},
  \bibinfo{author}{\bibfnamefont{V.}~\bibnamefont{Meden}}, \bibnamefont{and}
  \bibinfo{author}{\bibfnamefont{K.}~\bibnamefont{Sch\"onhammer}},
  \bibinfo{journal}{Rev. Mod. Phys.} \textbf{\bibinfo{volume}{84}},
  \bibinfo{pages}{299} (\bibinfo{year}{2012}).

\bibitem[{\citenamefont{Reuter and Wetterich}(1994)}]{Reuter94b}
\bibinfo{author}{\bibfnamefont{M.}~\bibnamefont{Reuter}} \bibnamefont{and}
  \bibinfo{author}{\bibfnamefont{C.}~\bibnamefont{Wetterich}},
  \bibinfo{journal}{Nucl. Phys. B} \textbf{\bibinfo{volume}{427}},
  \bibinfo{pages}{291} (\bibinfo{year}{1994}).

\bibitem[{\citenamefont{Tetradis}(1997)}]{Tetradis97}
\bibinfo{author}{\bibfnamefont{N.}~\bibnamefont{Tetradis}},
  \bibinfo{journal}{Nucl. Phys. B} \textbf{\bibinfo{volume}{488}},
  \bibinfo{pages}{92 } (\bibinfo{year}{1997}).

\bibitem[{\citenamefont{Sachdev}(2012)}]{Sachdev12}
\bibinfo{author}{\bibfnamefont{S.}~\bibnamefont{Sachdev}},
  \bibinfo{journal}{Phys. Rev. D} \textbf{\bibinfo{volume}{86}},
  \bibinfo{pages}{126003} (\bibinfo{year}{2012}).

\bibitem{Footnote}
For a lattice realization of the NCCP$^1$ model in $D=3$, 
$\nu = 0.75$ -- $0.8$ and $\eta_N = 0.2$ -- $0.4$ were obtained in Ref.~\onlinecite{Motrunich08}. 
Quantum Monte Carlo simulations of the two-dimensional $t$--$Q$ model report 
for the N\'eel to VBS transition
$\nu = 0.78(3)$ and $\eta_N = 0.26(3)$ in Ref.~\onlinecite{Sandvik07}, 
$\nu = 0.68(4)$ and $\eta_N = 0.35(3)$ in Ref.~\onlinecite{Melko08},
and $\nu = 0.51(3)$, and $\eta_N = 0.30(5)$ in Ref.~\onlinecite{Pujari13}.
Coming from the VBS side, Ref.~\onlinecite{Pujari13} also quotes 
$\nu_\mathrm{VBS} = 0.55(4)$ and $\eta_\mathrm{VBS} = 0.28(8)$.
%
%
%

\bibitem[{\citenamefont{Litim}(2001)}]{Litim01}
\bibinfo{author}{\bibfnamefont{D.~F.} \bibnamefont{Litim}},
  \bibinfo{journal}{Phys. Rev. D} \textbf{\bibinfo{volume}{64}},
  \bibinfo{pages}{105007} (\bibinfo{year}{2001}).

\end{thebibliography}


\end{document}